\documentclass[journal]{IEEEtran}
\IEEEoverridecommandlockouts
\usepackage{amsmath,amsfonts}
\usepackage{algorithmic}
\usepackage{array}
\usepackage{tabularx}
\usepackage{verbatim}
\usepackage{graphicx}
\usepackage[caption=false]{subfig}
\usepackage{multirow}
\usepackage{placeins}
\usepackage{xcolor}
\usepackage[switch]{lineno}
\begin{document}
\title{Passive Acoustic–based Composite Indices for Reef Health Monitoring in Noisy Tropical waters}
\author{Hari Vishnu\IEEEauthorrefmark{1}\IEEEauthorrefmark{4},~\IEEEmembership{Senior Member,~IEEE,} Yuen Min Too\IEEEauthorrefmark{1}\IEEEauthorrefmark{4},~\IEEEmembership{Member,~IEEE,} Mandar Chitre\IEEEauthorrefmark{1}\IEEEauthorrefmark{2},~\IEEEmembership{Fellow,~IEEE,} Danwei Huang\IEEEauthorrefmark{3}, Teong Beng Koay\IEEEauthorrefmark{1},~\IEEEmembership{Senior Member,~IEEE,} Sudhanshi S. Jain\IEEEauthorrefmark{5}

\IEEEauthorblockA{\IEEEauthorrefmark{1}\textit{Acoustic Research Laboratory, Tropical Marine Science Institute, National University of Singapore}}

\IEEEauthorblockA{\IEEEauthorrefmark{2}\textit{Department of Electrical \& Computer Engineering, National University of Singapore}}

\IEEEauthorblockA{\IEEEauthorrefmark{3}\textit{Lee Kong Chian Natural History Museum, National University of Singapore}}

\IEEEauthorblockA{\IEEEauthorrefmark{5}\textit{Department of Biological Sciences, National University of Singapore}}

}

\maketitle

\def\thefootnote{\IEEEauthorrefmark{4}}\footnotetext{These authors contributed equally to this work}

\begin{abstract}
Passive acoustic monitoring offers the potential to enable long-term, spatially extensive assessments of coral reefs. To explore this approach, we deployed underwater acoustic recorders at ten coral reef sites around Singapore waters over two years. To mitigate the persistent anthropogenic and current-induced noise masking the low-frequency reef soundscape, we trained a convolutional neural network denoiser. Analysis of the acoustic data reveals distinct morning and evening choruses. Though the correlation with environmental variates was obscured in the low-frequency part of the noisy recordings, the denoised data showed correlations of acoustic activity indices such as sound pressure level and acoustic complexity index with diver-based assessments of reef health such as live coral richness and cover, and algal cover. Furthermore, the shrimp snap rate, computed from the high-frequency acoustic band, is robustly correlated with the reef parameters, both temporally and spatially. This study demonstrates that passive acoustics holds valuable information that can help with reef monitoring, provided the data is effectively denoised and interpreted. This methodology can be extended to other marine environments where acoustic monitoring is hindered by persistent noise.
\end{abstract}


\section{Introduction}
Coral reefs provide complex and varied habitats that support approximately 25\% of all marine biodiversity on the planet. Millions of people in tropical regions depend on reefs for food, protection, and employment. In recent years, climate change, pollution, unsustainable fishing practices and destructive coastal development have destroyed more than 60\% of reefs worldwide~\cite{burke2011reefs}. This study focuses on reefs in tropical shallow water regions, with Singapore serving as a representative example. Singapore, as one of the world’s busiest trans-shipment hubs, is not spared of anthropogenic impacts on reef health, having lost up to 65\% of its live coral cover since 1986~\cite{chou2000southeast}. Hence, the importance of safeguarding the remaining coral reefs cannot be overstated. While tremendous efforts have been undertaken in restoration (eg.~\cite{ng2018coral,jaap2000coral} in Singapore waters), effective conservation needs long-term and regular monitoring to track the growth, ecological status and recovery. Existing monitoring methods rely mainly on visual surveys, either through diver-based surveys or video assessments which are often manpower intensive, expensive, capable of covering only short segments of space and time, and especially difficult to perform in the turbid waters around Singapore~\cite{chou1996response}. 

Healthy reefs are characterized by vibrant soundscapes owing to symphonies orchestrated by diverse soniferous organisms such as snapping shrimp and fish, as well as ecological processes like spawning and feeding~\cite{chitre2012snapping,lobel1992sounds}. In contrast, degraded reefs are notably quieter and less attractive to marine fauna, reflecting ecosystem decline. This contrast opens up the feasibility of using passive acoustic monitoring (PAM) to monitor reef health and associated parameters. Because sound travels efficiently over long distances underwater, PAM can capture rich environmental information facilitating monitoring over large spatial scales. Furthermore, PAM is non-invasive, is unhindered by bad visibility and time of the day, and provides good temporal resolution. PAM has been successfully applied for monitoring in a wide range of ecosystems ranging from polar~\cite{Kinda2013} to subtropical~\cite{Lin2021} to tropical ones~\cite{nedelec2015soundscapes}, and through applications as diverse as defense, infrastructure monitoring and marine mammal monitoring~\cite{Fleishman2023}. Reflecting its growing importance, sound has been incorporated as an essential ocean variable in the global ocean observing system~\cite{Tyack2023}. Recent advancements in compact acoustic recording systems have further increased interest in studying the acoustic signatures of healthy and degraded reefs. These relatively inexpensive systems (a few thousand USD) can continuously operate for extended periods, on the order of months. This enables collection of long-term spatio-temporal data crucial for understanding patterns and trends in reef health that may not have been immediately apparent in shorter-term studies~\cite{lammers2008ecological,mcwilliam2018coral}. 

Accumulating evidence shows that reef soundscapes are closely linked to reef health. The strength of diel and spatial trends of low frequency sound, usually dominated by fish calls, is linked to coral cover and fish density~\cite{kaplan2015coral,nedelec2015soundscapes, Kennedy2010, duarte}. Acoustic indices exhibit positive correlations with density of benthic invertebrates and reef species diversity~\cite{freeman2016rapidly,harris2016ecoacoustic,bertucci2016acoustic, McCammon2025, Lin2023}, and the acoustic activity of benthic invertebrates such as snapping shrimp has been used to monitor post-bleaching reef recovery \cite{Raick2025}. 
Furthermore, the soundscape may also play an ecological role, guiding the larvae and juveniles of some reef species as a navigational cue to locate settlement habitats, even facilitating restoration \cite{Hodson2025}. Some works have tapped into the power of machine learning to discriminate healthy and degraded reef sites by combining ecoacoustic indices~\cite{williams2022enhancing}, whereas others have aimed to summarize the soundscapes in terms of dynamical complexity~\cite{Siddagangaiah2022}. An important finding of these works is that multi-index approaches outperform single metrics in isolation, in terms of robustly characterizing biodiversity~\cite{Sueur2014}. Clustering and source separation approaches have also been used to separate the biotic and abiotic components of soundscapes to focus on reef-produced sound~\cite{Lin2021}. While acoustic indices could be used to assess ecological states of coral reefs, they are susceptible to masking by non-biological noises such as anthropogenic and geophysical noise, which have to be segregated from the data to mitigate their effect~\cite{lin2017improving,elise2019optimised}. This is crucial for ensuring the accuracy and reliability of acoustic-based reef health assessments, especially in shallow-water regions like Singapore with heavy shipping and biological activity, where anthropogenic and other natural sources of noise can obscure use of acoustics for biodiversity assessment in certain frequency bands~\cite{Vishnu2024}. This underscores the importance of noise mitigation strategies in underwater soundscape analysis.

In this study, we conducted coral reef site surveys for acoustic and transect data collection, as detailed in Section~\ref{section:surveys}. We analyze the acoustic recordings and identify persistent noise sources that impede the acoustic assessment of reef activity in Section~\ref{section:data}. To address this issue, we develop a machine-learning based `Reef denoiser' to remove noise from reef-dominated biological soundscapes utilizing supervised deep learning, outlined and assessed in Section~\ref{section:denoising}. Subsequently, the denoiser is employed on the acoustic data, and in Section~\ref{section:pam}, indices derived from the data are shown to correlate well with direct environmental measures of reef health in Singapore waters. Following this, a composite acoustic index combining multiple indices is developed for effective reef health monitoring. Our findings highlight the feasibility of  passive acoustic monitoring for long-term large-scale assessment of reef-health, even in noisy waters such as those found in Singapore.

\begin{figure}[tbp!]
\centering
 \subfloat[Position of sensors\label{fig:recorder:position}]{
       \includegraphics[width=0.6\linewidth]{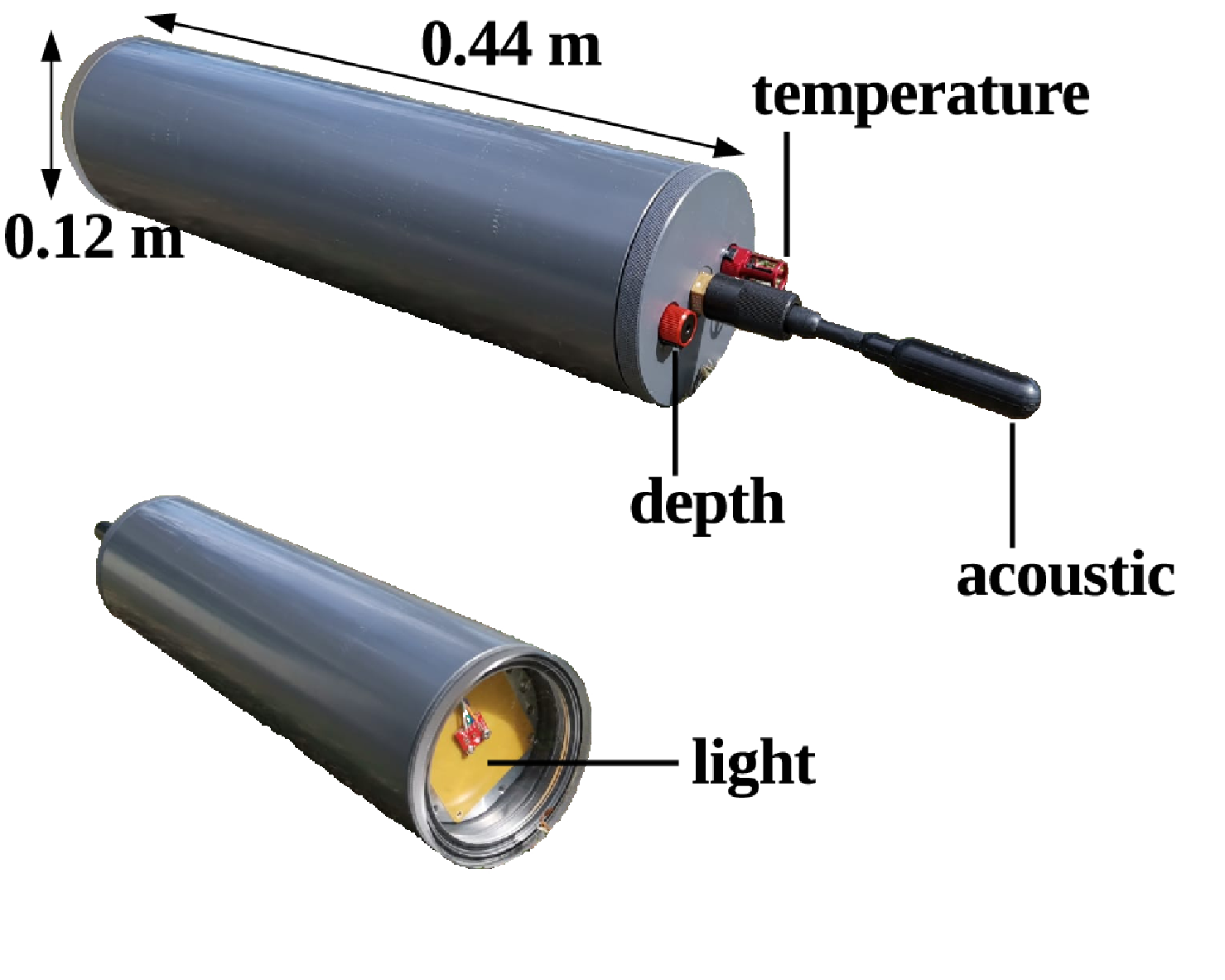}}
    \hfill
\centering
  \subfloat[Anchoring bracket\label{fig:recorder-fixture}]{
        \includegraphics[width=0.26\linewidth]{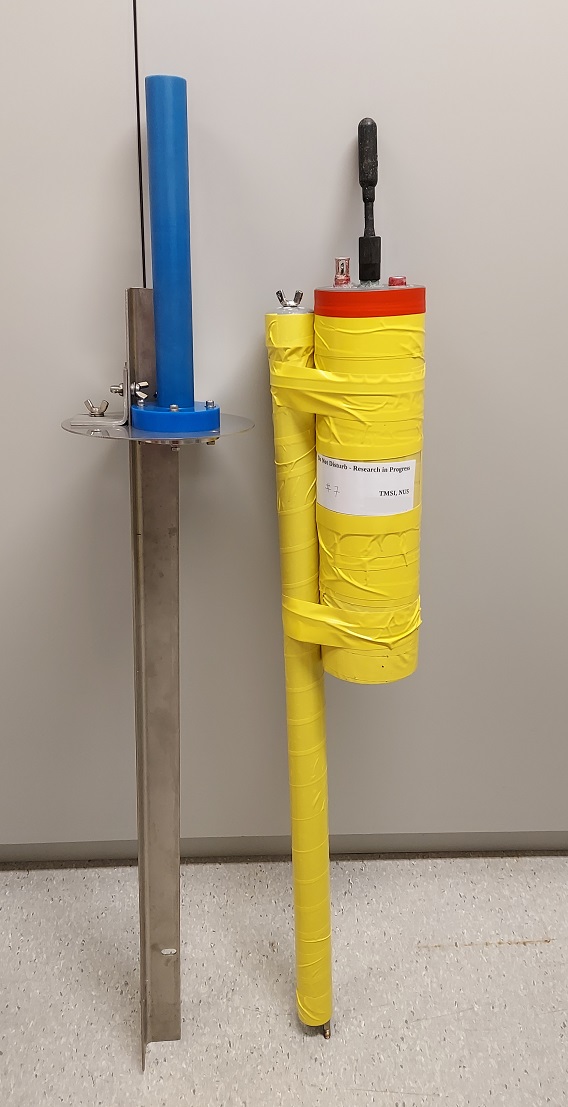}}
\caption{Passive recorder and anchoring bracket for deployment. The blue bar is a mounting structure for the PVC pipe to slot in.}
\label{fig:recorder}
\end{figure}

\begin{table}[tbp!]
\renewcommand{\arraystretch}{1.3}
\caption{Deployment location and amount of recorded data at different sites.}
\label{table:sites_days}
\centering
\begin{tabular}{|p{3cm}|p{1.7cm}|p{3.0cm}|}
\hline
Site & Amount of data (days) & Location \\
\hline\hline
Hantu & 231.85 & $1^{\circ}13.62^{\prime}\text{N}$,$103^{\circ}44.80^{\prime}\text{E}$ \\
\hline
Jong & 156.49 & $1^{\circ}12.83^{\prime}\text{N}$,$103^{\circ}47.25^{\prime}\text{E}$\\
\hline
Kusu & 128.90 & $1^{\circ}13.54^{\prime}\text{N}$,$103^{\circ}51.59^{\prime}\text{E}$\\
\hline
Raffles lighthouse & 158.14  & $1^{\circ}9.63^{\prime}\text{N}$,$103^{\circ}44.42^{\prime}\text{E}$\\
\hline
Subar Darat (Big Sisters) & 191.73 & $1^{\circ}13.00^{\prime}\text{N}$,$103^{\circ}49.88^{\prime}\text{E}$\\
\hline
Subar Laut (Little Sisters)& 116.57 & $1^{\circ}12.75^{\prime}\text{N}$,$103^{\circ}50.17^{\prime}\text{E}$\\
\hline
Semakau-Northwest & 132.98 & $1^{\circ}12.69^{\prime}\text{N}$,$103^{\circ}45.36^{\prime}\text{E}$\\
\hline
Semakau-Southwest & 66.88  & $1^{\circ}12.13^{\prime}\text{N}$,$103^{\circ}45.31^{\prime}\text{E}$\\
\hline
Seringat & 204.38 & $1^{\circ}13.71^{\prime}\text{N}$,$103^{\circ}51.34^{\prime}\text{E}$\\
\hline
Terumbu Pempang Tengah (TPT) & 170.76  & $1^{\circ}13.62^{\prime}\text{N}$,$103^{\circ}43.66^{\prime}\text{E}$\\
\hline
\end{tabular}
\end{table}
\section{Data collection} \label{section:surveys}
\subsection{Instrumentation}
We used a customized version of Loggerhead LS1 recorders\footnote{https://www.loggerhead.com/ls1-ls2-recorders} to collect ambient underwater noise in sites with current or historical coral reef cover. Each recorder is equipped with a HTI-96-Min hydrophone, a Bar02 depth sensor, a Celsius temperature sensor and a SparkFun RGB light sensor. The hydrophone has a flat frequency response over the $0.002-30$~kHz range. The depth sensor can measure up to $10$~m depth with a resolution of $0.16$~mm. 

The recorder is compact, measuring $44$~cm in length and weighing just a few kgs, facilitating easy deployment. Designed for low-power operations, the device, powered by $12$ D-cell batteries, can continuously acquire data for $\sim40$ days. To secure the recorder onto the seabed at a reef site, an anchoring bracket was developed. It consists of one stainless steel L-bar and one PVC pipe. During deployment, the L-bar is anchored in the sea floor to function as a base while the PVC pipe is used to mount the recorder. Fig.~\ref{fig:recorder} illustrates the recorder and the anchoring bracket.

\subsection{Acoustic recorder deployment}
\begin{figure}[tbp!]
\centering
\includegraphics[width=0.9\linewidth]{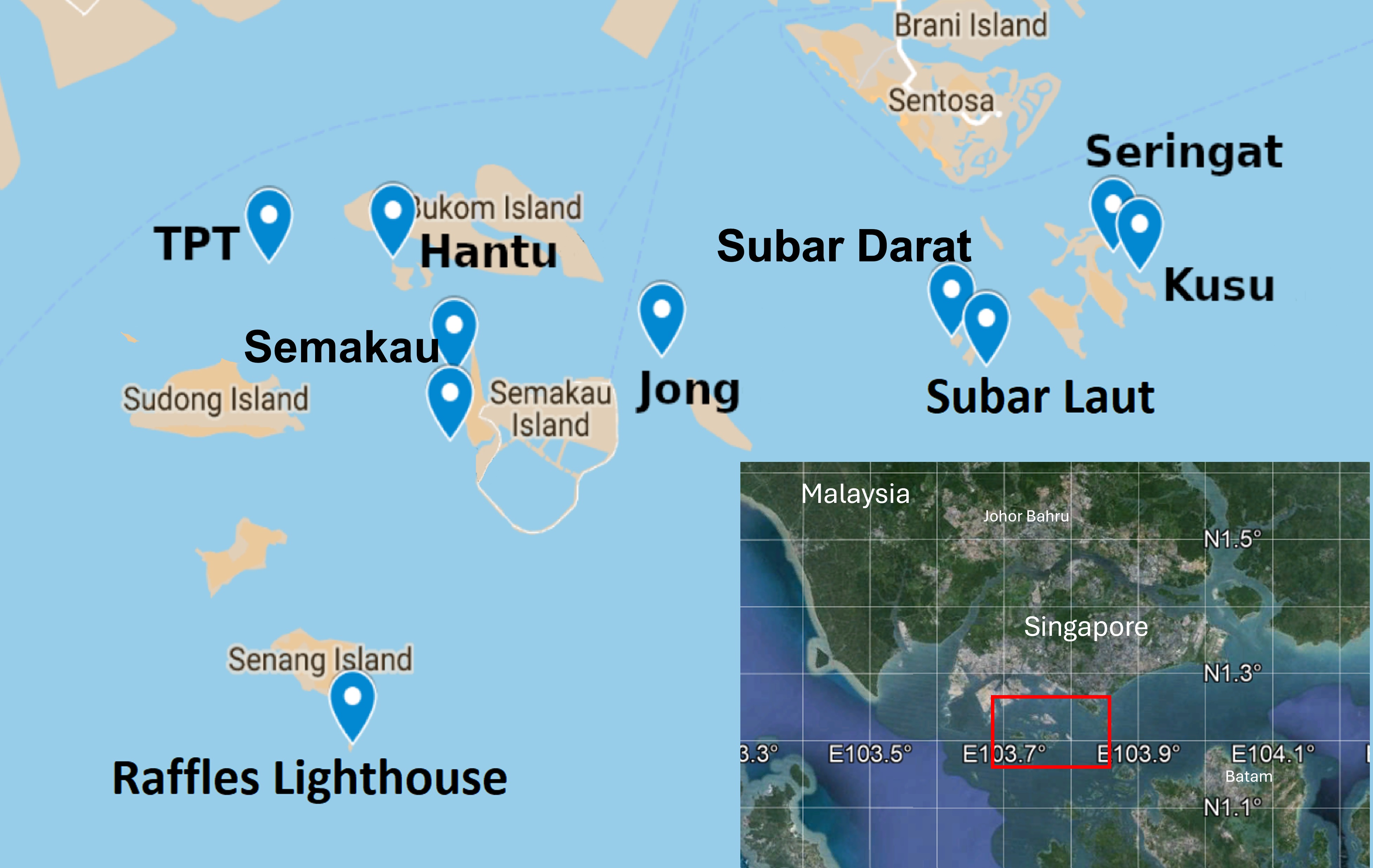}
\caption{Locations in Singapore strait near reefs where passive acoustic recorders were deployed (blue markers), and (inset) zoomed out map of the region around Singapore showing the area studied marked by a red box.}
\label{fig:googlemap}
\end{figure}

\begin{figure}[tbp!]
\centering
\subfloat[Preparing for the deployment.\label{fig:deployment_photo_1}]{%
    \includegraphics[width=0.24\textwidth]{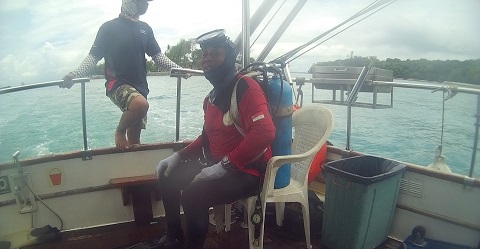}%
}%
\subfloat[Searching for the location.\label{fig:deployment_photo_2}]{%
    \includegraphics[width=0.24\textwidth]{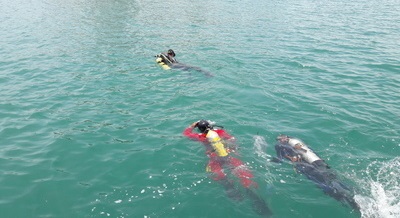}%
}%
\\
\subfloat[Approaching the location to install the recorder.\label{fig:deployment_photo_3}]{%
    \includegraphics[width=0.24\textwidth]{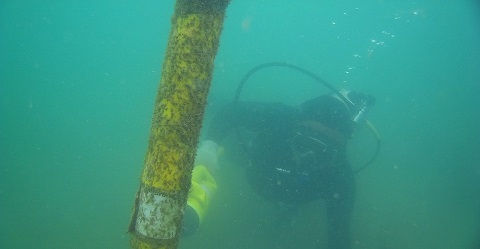}%
}%
\subfloat[A PVC pole was used as a marker when no recording was done.\label{fig:deployment_photo_4}]{%
    \includegraphics[width=0.24\textwidth]{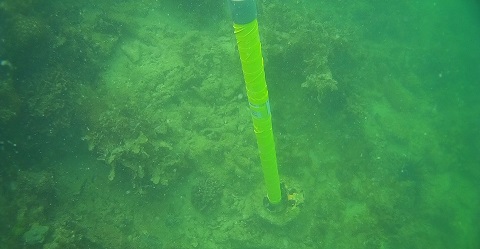}%
}%
\\
\subfloat[A recorder was installed after removing the marker.\label{fig:deployment_photo_5}]{%
    \includegraphics[width=0.24\textwidth]{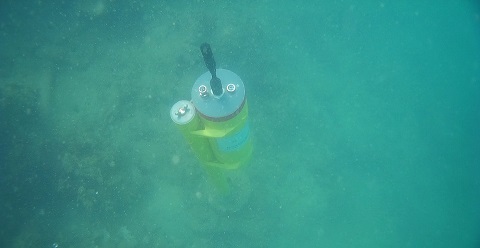}%
}%
\subfloat[Recorders after deploying for one month.\label{fig:deployment_photo_6}]{%
    \includegraphics[width=0.24\textwidth]{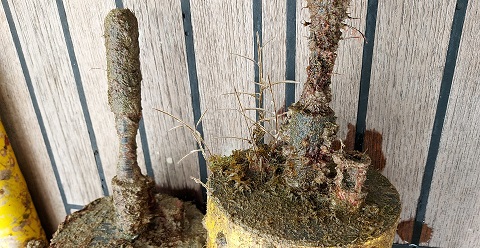}%
}%
\caption{Photos taken during the deployments.}
\label{fig:deployment_photos}
\end{figure}

Ten sites were identified for ambient noise data collection within shallow water coral reef areas, listed in Table~\ref{table:sites_days} along with the duration of data-collection at each site. The site locations, labeled in Fig.~\ref{fig:googlemap}, cover a large part of the Singapore Strait. 
For each deployment, two divers installed the anchoring bracket and secured the passive recorder into the mounting structure. We completed most of the deployments over $\sim$2 years, starting from June 2019 to August 2021. Each site had $3-5$ deployments, with each deployment lasting approximately 1 month. 
Photos in Fig.~\ref{fig:deployment_photos} depict the deployment procedure. Deployments scheduled between April-June 2020 were interrupted due to COVID-19 pandemic response measures in Singapore. 
The recorder captured continuous acoustic pressure data in the form of 5-minute WAV files, each storing one channel audio sampled at $96$~kHz with $16$-bit resolution. 

\subsection{Coral reef visual survey}
Visual observations of hard coral cover, algal cover and fish abundance serve as key indicators for reporting coral reef health~\cite{obura2019coral}. Hard coral cover and composition determine the reef structure, providing critical habitat for many organisms. Algal cover consists of different algae groups serving unique functional roles in reef communities. Macroalgae are primary producers which provide habitats for numerous reef-associated species \cite{Low2019}. 
Transect surveys to assess these parameters were conducted at the selected coral reef sites during 2019-2021. Five $20$~m transects were repeated along the reef crest (2-3~m depth below chart datum) spaced 3-5~m apart during each survey \cite{Chan2024}, and 1-3 surveys were conducted for each site throughout these 2 years. Line intercept transect observations were performed, during which all benthic components directly under the transect line were recorded and quantified for cover. Measured cover and computed indices were averaged across the five transect replicates for each survey. Table~\ref{table:transect} provides a summary of the variables measured in the transect data. Note that some of the variables overlap or are hierarchical (e.g., macroalgal cover is a subset of total algal cover).
\begin{figure}[tbp!]
    \includegraphics[width=1.05\linewidth]{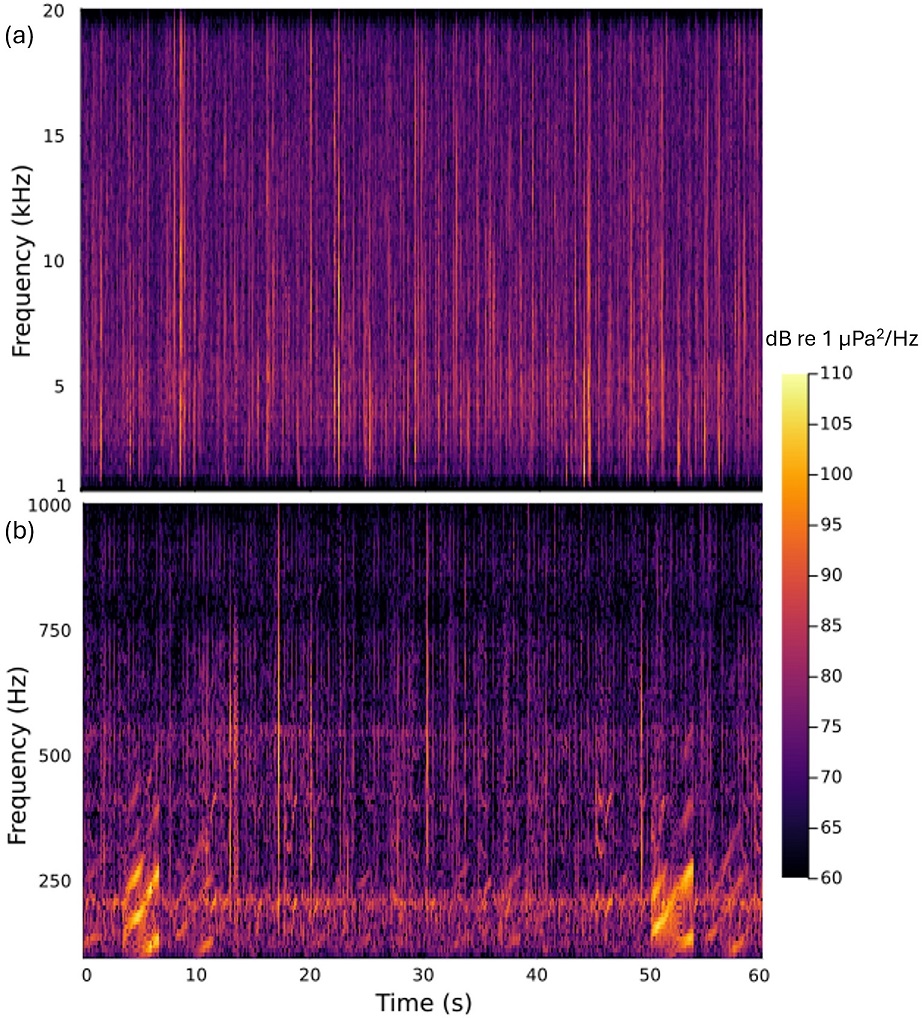}
\caption{Spectrograms depicting 1 minute of acoustic data in two different bands, revealing the presence of (a) shrimp snaps in the 1-20~kHz band, and (b) fish chorus in the 0.1-1~kHz band.}
\label{fig:spectrograms}
\end{figure}
\begin{figure}[tbp!]
    \centering
    \includegraphics[width=1.05\linewidth]{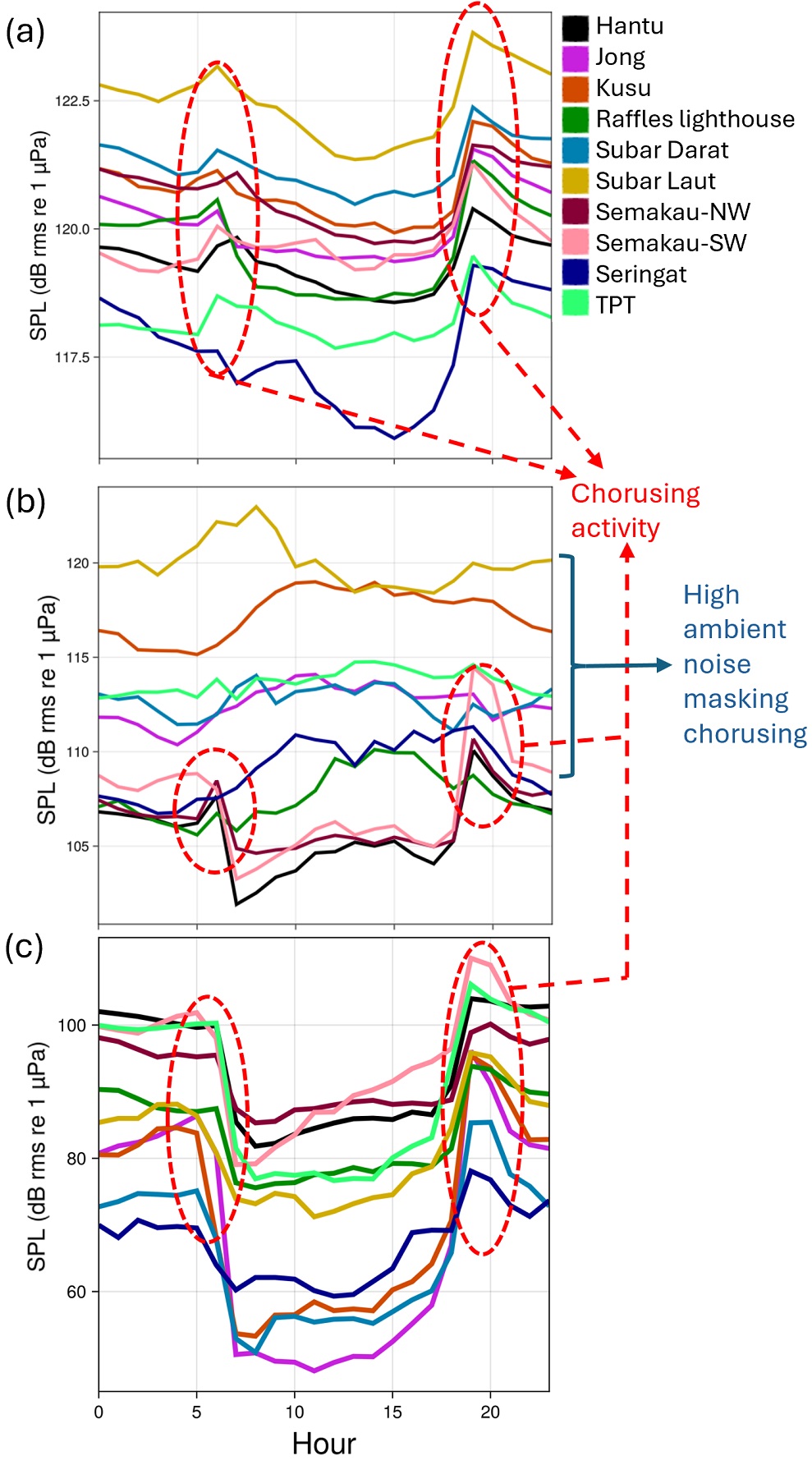}
\caption{The average SPL over a 24-hour period (a) in the high-frequency 1-20~kHz band, (b) in the low-frequency 0.1-1~kHz band without denoising, and (c) in the low-frequency band after denoising using the Conv-TasNet Reef denoiser. While the choruses are visible in (a) in the form of peaks around 4-7~AM and 6-9~PM, they are not apparent for many of the sites in (b). Also, (b) shows a much larger vertical spread between SPL at different sites due to the impact of additional noise sources. Applying the denoiser reveals the choruses in the low-frequency band in (c), indicating that the noise from shipping and flow effects is effectively suppressed, allowing better observation of biologically relevant acoustic activity.}
\label{fig:spl}
\end{figure}
\begin{table}[tbp!]
\caption{Descriptions of transect variables.}
\label{table:transect}
\centering
\begin{tabularx}{\linewidth}{|l|X|}
\hline
Variable & Description \\
\hline\hline
Live Coral Richness & number of live coral species \\
\hline
Live Coral Size & mean diameter of live coral colonies \\
\hline
Live Coral Cover & percentage cover of live coral colonies \\
\hline
Dead Coral Cover & percentage cover of dead coral including freshly dead and algae-covered \\
\hline
Invertebrate Cover & percentage cover of sessile organisms, including sponge, soft coral, zoanthid and others \\
\hline
Algal Cover & percentage cover of all algae including macro, turf, assemblage and coralline algae \\
\hline
Macroalgal Cover & percentage cover of large canopy-forming macroalgae \\
\hline
\end{tabularx}
\end{table}

\section{Acoustic data} \label{section:data}
\subsection{Frequency bands and bio-acoustic sources}
The acoustic recordings are rich in snapping shrimp crackles and fish vocalizations, with occasional marine mammal vocalizations also picked up~\cite{Vishnu2024}. The snapping shrimp crackle is an ensemble of many transient impulsive signals, known as snaps, resulting from the collapse of cavitation bubbles generated by rapid shrimp claw closure~\cite{versluis2000snapping}. These broadband snap signals dominate the frequencies ranging from 2~kHz to over 200~kHz \cite{Potter1997, potter1997acoustic, Vishnu2024}. Fig.~\ref{fig:spectrograms}(a) shows a spectrogram of sample acoustic data recorded during one of the evenings, focusing on the 1-20~kHz band, filled with snaps (visible as vertical stripes). 

Fish vocalizations exhibit a wide range of types, including whistling, honking, drumming, and clicking \cite{Fishsounds, looby2023fishsounds}. Majority of fish-sounds are typically within the 0.1-1~kHz frequency band~\cite{kaplan2015coral,kaplan2018acoustic,McCammon2025}. Fig.~\ref{fig:spectrograms}(b) shows a spectrogram of the same recording as (a), but focusing on the 0.1-1~kHz band, highlighting fish vocalizations (frequency-modulated signals).

In the forthcoming analysis, the 0.1-1~kHz and 1-48~kHz bands, which are dominated by fish and snapping shrimp activity respectively, are denoted as low and high-frequency bands respectively, and are analyzed separately. After bandpass filtering into these two bands, the sound pressure level (SPL) was calculated and averaged over 1-minute segments. Fig.~\ref{fig:spl}(a) and~(b) show the daily SPL variation for different sites and for the high and low-frequency bands, averaged over the entire recording period, summarizing the diel and spatio-temporal variability in the soundscapes. 
\begin{figure}[tbp!]
\includegraphics[width=1.0\linewidth]{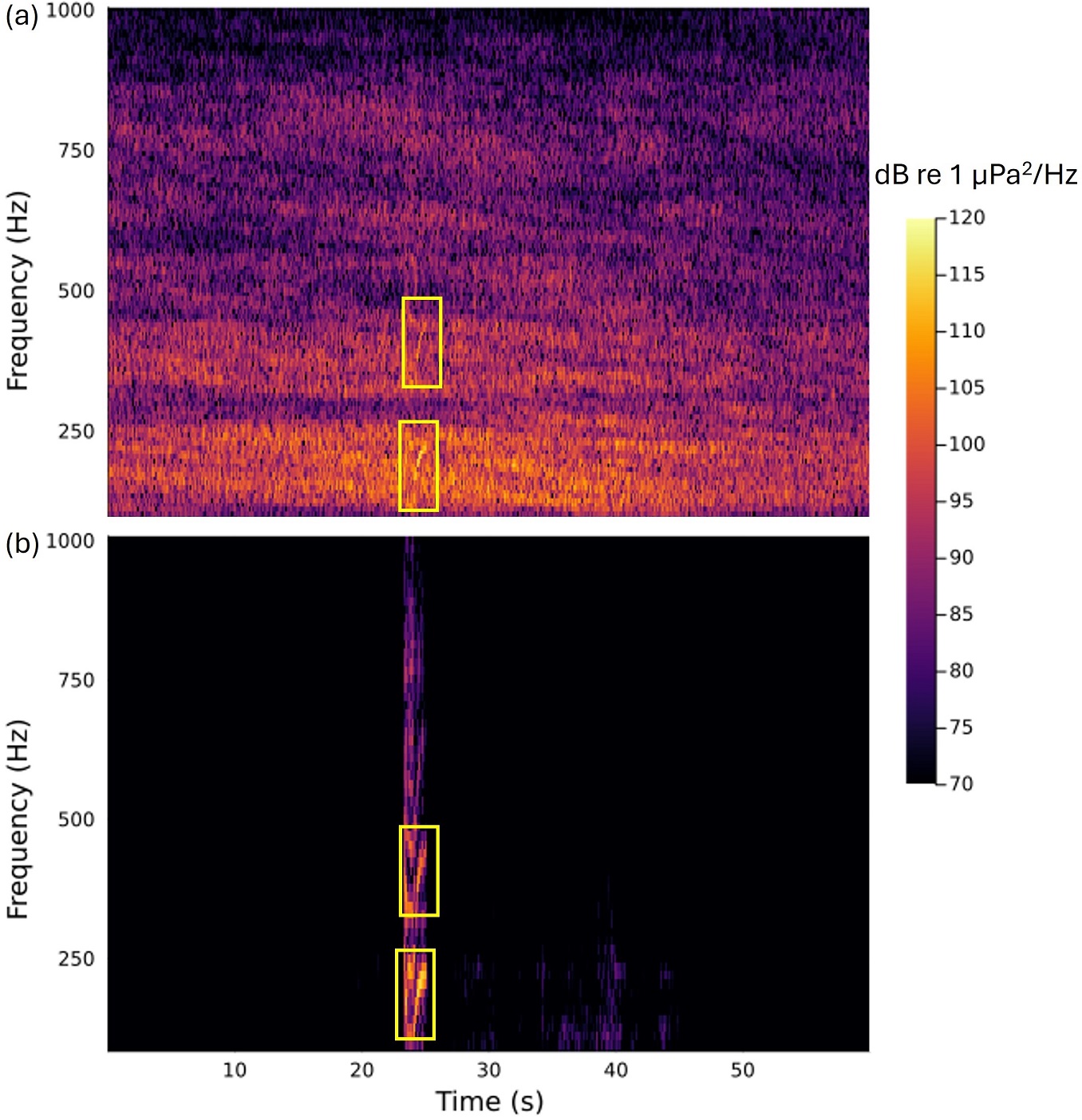}
\caption{Spectrograms illustrating one minute of (a) raw acoustic data within the 0.1-1~kHz band showing vessel-generated noise masking a fish vocalization, and (b) acoustic data denoised by the ConvTasNet Reef denoiser, revealing the vocalization.}
\label{fig:spectrograms:shipnoise}
\end{figure}

\subsection{Case of the missing choruses}
Heightened activity of marine organisms may often be observed during the early morning and late evening hours, and is referred to as chorusing. In Fig.~\ref{fig:spl}(a), a notable feature is that distinct peaks emerge during the 4-7~AM and 6-9~PM periods at all the sites, corresponding to the morning and evening snapping shrimp choruses, as widely reported in the literature \cite{JOHNSON1947, Potter1997}. 

The low-frequency data in Fig.~\ref{fig:spl}(b), in a dramatic turn of events, exhibits this double-peak pattern only at Hantu, Semakau-NW, and Semakau-SW; it is notably absent at other sites. So why are these choruses missing? 

There are two possible explanations: either a paucity of marine life at these sites, or masking of the low-frequency data by other noise sources. As visible in Fig.~\ref{fig:spl}(b), the low-frequency SPL varies by as much as $\sim25$~dB, i.e., more than two orders of magnitude,  a considerably broader range compared to the $\sim$6~dB spread observed in the high-frequency data, suggesting that masking by extraneous noise is the more plausible explanation for the lack of consistent patterns across the sites. This is further confirmed by the observation that the sites with lowest recorded low-frequency SPLs (Hantu and Semakau) do exhibit the chorus pattern, indicating that the chorusing is masked in the remaining sites due to higher noise floor resulting from extraneous noise contamination. Hence, we hypothesize that non-biological sounds are responsible for the wide spatial SPL variation, and are masking the biological choruses in the low-frequency band.

\begin{figure}[tbp!]
\includegraphics[width=0.9\linewidth]{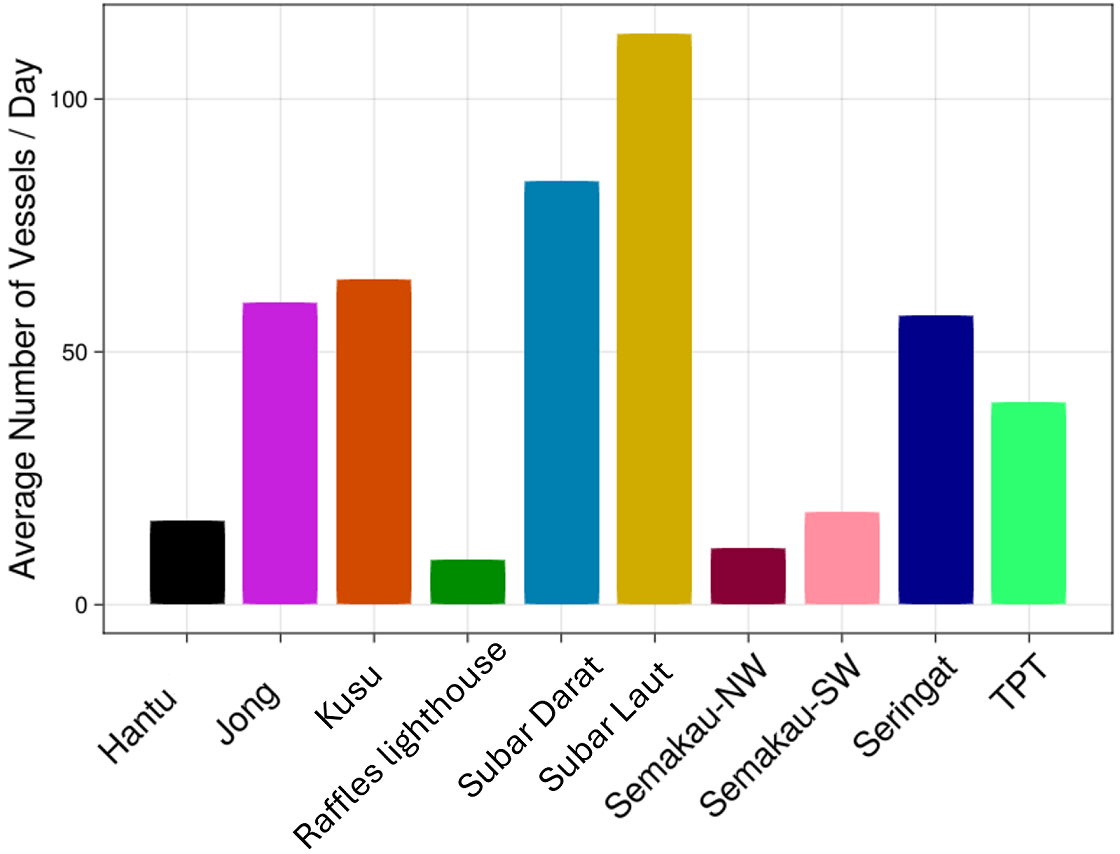}
\caption{The daily average count of marine vessels within a 1-kilometer radius of each site during two months in end-2021.}
\label{fig:ais}
\end{figure}
\subsection{The suspects \textemdash shipping and tidal-induced noise}
As Singapore is a major global trans-shipment hub, shipping noise is a major source of interference in the low-frequency band. In recent years, the number of marine vessels globally has increased significantly (a factor of 4 from 1992-2012), leading to a corresponding increase in shipping noise, with low-frequency ambient noise levels rising as fast as 3~dB/decade \cite{Erbe2019}. Source levels can vary from 130-160~dB~re~1~$\mu$Pa for small craft, to upto 200~dB or more for larger or faster ships~\cite{Erbe2019}. The frequency range of the recorded shipping noise varies depending on factors such as the type of ship, its speed, and the distance between the ship and the recorder \cite{Erbe2019}. Typically, it covers a range of frequencies from 10~Hz to 1~kHz~\cite{potter1997acoustic, Potter1997}. Fig.~\ref{fig:spectrograms:shipnoise}(a) presents a 1-minute spectrogram showing boat noise dominating the low-frequency band and obscuring a fish vocalization present in the data between 20-30~s. This exemplifies the substantial noise contamination present at sites adjacent to anchorages and busy shipping lanes. Automatic Identification System (AIS) data presented in  Fig.~\ref{fig:ais} indicate that Hantu, Semakau Northwest, and Semakau Southwest experience lower marine traffic in their vicinity. This observation matches the SPL patterns in Fig.~\ref{fig:spl} and explains why biological choruses are observable in recordings at these sites but not at others. The absence of choruses at Raffles Lighthouse, despite low marine traffic, is likely attributable to tidal flow-induced noise, discussed next.

\begin{figure}[tbp!]
\includegraphics[width=1.0\linewidth]{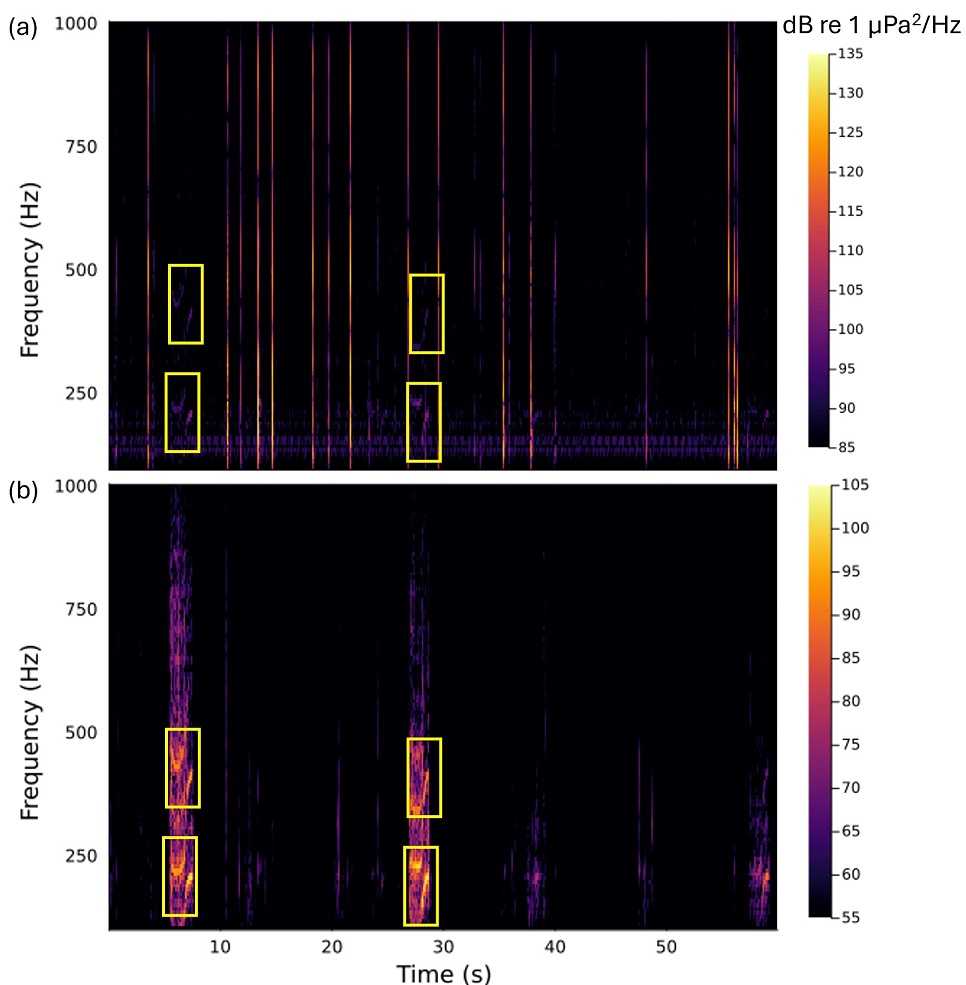}
\caption{Spectrogram displaying one minute of (a) raw acoustic
data within the 0.1-1 kHz band showing tidal flow-induced noise
masking four fish vocalizations, and (b) acoustic data denoised by
ConvTasNet, revealing the fish vocalizations.}
\label{fig:spectrograms:flownoise}
\end{figure}

The Singapore Strait is characterized by highly dynamic and variable currents, primarily attributed to its location between the Indian Ocean to the West and the Pacific Ocean to the East. This exposes the strait to strong monsoon-driven flows with velocities of up to 3-4 knots~\cite{tidetables2021}. Sites such as Raffles lighthouse and Subar Laut, which lie near the main channel of the Singapore Strait, are particularly impacted by these potent tidal currents. Fig.~\ref{fig:spectrograms:flownoise}(a) provides a spectrogram of a 1-minute recording at Raffles lighthouse dominated by tidal-current induced noise, with vertical stripes corresponding to knocking sounds, likely due to vortex-induced vibrations. 

\begin{figure*}[tbp!]
\centering
    \includegraphics[width=0.85\textwidth]{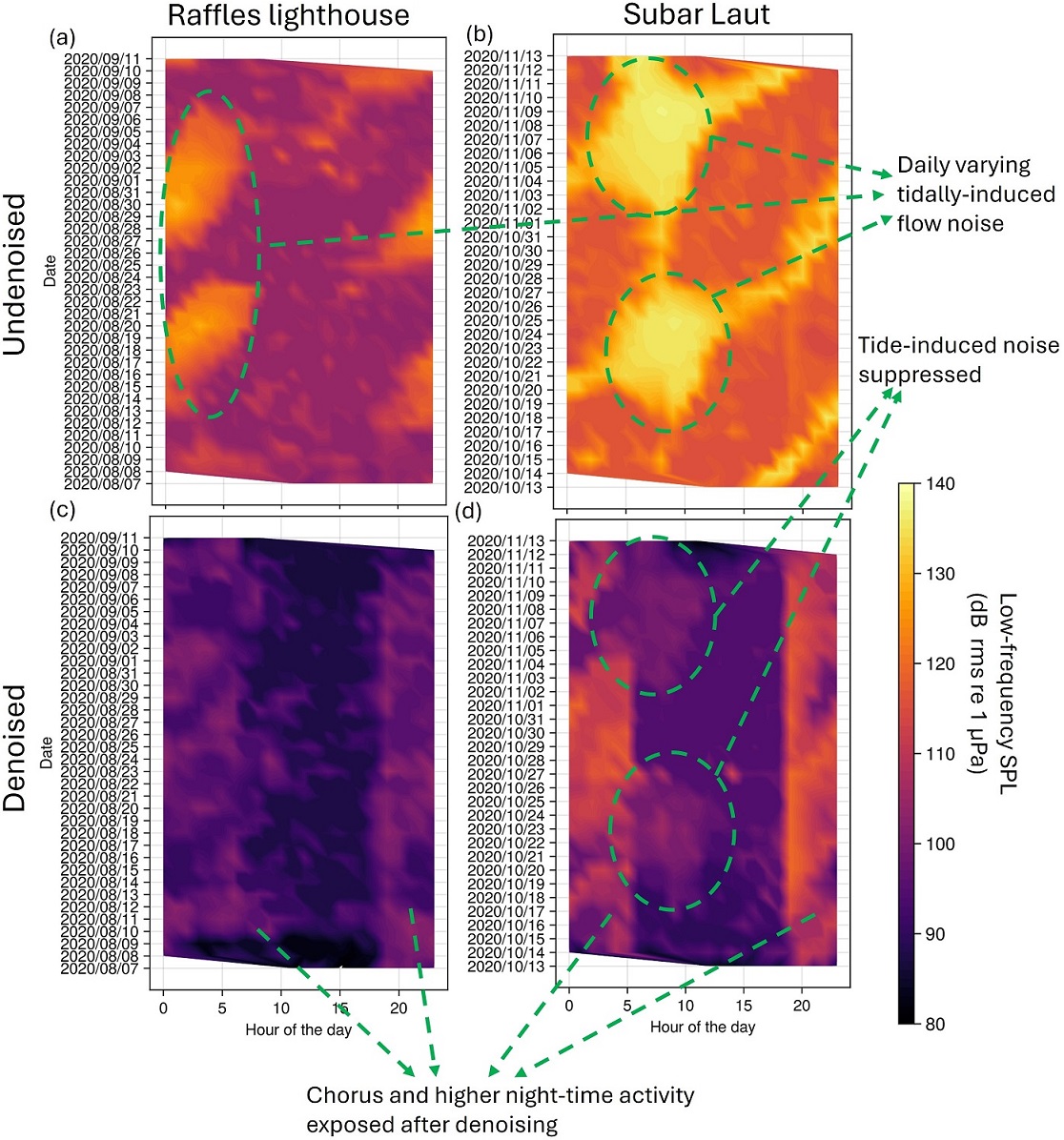}
\caption{Daily variation of 1-minute average of low-frequency SPL across various dates. 
}
\label{fig:datehour}
\end{figure*}

To further elucidate the influence of these water flows on the acoustic environments of Raffles lighthouse and Subar Laut, Fig.~\ref{fig:datehour}(a) and (b) show heatmaps of SPL recorded at different dates and times at these sites. Each heatmap covers $\sim$1 month of acoustic data per deployment. Certain deployments exhibit recurring loud events (diagonal yellow streaks), lasting about five hours each day, shifting by an hour daily and spanning a ten day duration, and occuring twice monthly. This pattern aligns with tidal cycles, confirming that the soundscapes of these sites were strongly influenced by tidal current-induced noise. Additional horizontal patches in the heatmaps indicate periods of ship noise during the deployments.

The high-frequency acoustic recordings, primarily dominated by shrimp snaps, are notably cleaner in quality when contrasted with the low-frequency recordings. The intrusion of additional noise, stemming from ship and current-related sources, has effectively masked the lower-frequency biological sounds originating from the coral reefs. The pervasiveness of such disruptive noises has the potential to impede data analysis for passive acoustic monitoring. To overcome this challenge, we propose a machine learning-based method in the next section to mitigate noise contamination in the low-frequency biological soundscapes. 

\section{Denoising biological soundscapes} 
In order to facilitate acoustic reef health assessment from the noisy dataset recorded, one approach could be to focus on detecting specific fish calls of interest, as for example, done in \cite{McCammon2025}. However, in our scenario, we lack enough labeled samples for fish calls, and do not have an extensive database of all the different types of biodiversity-originating calls in Singapore waters. Furthermore, we find that the noise is a more delimiting factor in using the recordings effectively, and hence we do not focus on this method. 

A second approach would be to focus on removing known noise elements from the recordings. This approach would not require labeled samples of the fish vocalizations of interest, but it would require accurate characterization of the noise elements to be removed. In this section, we focus on this second approach because we have a lot of ambient noise data available from local waters obtained from past studies as well as online sources, that allows us to train and test an effective denoiser to enable acoustic assessment of reef health. A related approach to this is that used in \cite{Lin2021} by separating sound sources in the recordings, but in this work the authors use unsupervised methods for the separation, whereas we choose to develop a denoising technique in a supervised fashion using the available noise data.

\label{section:denoising}
\subsection{Development of denoiser for low-frequency band}
Heuristic soundscape denoising methods, including techniques like notch filtering, adaptive line enhancers~\cite{yan2006background}, and sound source separation~\cite{lin2017improving}, inherently depend on pre-existing knowledge or underlying assumptions about the characteristics of the signals and noises. However, in practical applications, securing this essential knowledge or formulating precise assumptions can prove to be a formidable challenge, especially when dealing with signals and noises that exhibit extensive variability. In tropical waters, denoisers to suppress shrimp sound have been explored in previous studies~\cite{Vishnu2024, Mahmood2017}, but these are not applicable in the current work where it is of interest to retain the shrimp sound as a biological signal of interest to utilize. In this study, we wish to mitigate shipping noise in the low-frequency band, which exhibits a wide range of variability and is not easy to characterize statistically in a way that simple linear denoising filters can tackle. Furthermore, we also wish to mitigate the effect of flow-induced noise, which has very different characteristics from shipping noise. 

To tackle these different kinds of noises in a manner specifically tuned to our application area of interest (Singapore waters),  we explore three deep learning architectures originally developed for audio source separation and adapt them to the task of soundscape denoising, namely (1) Conv-TasNet~\cite{luo2019conv}, (2) WaveUNet~\cite{Stoller2018} and (3) DEMUCSv2~\cite{Defossez2021}. Our premise for using these approaches is that audio denoising can be seen as a specialized instance of the broader challenge of audio separation, in which the goal is to recover a single target signal from a noisy mixture. Unlike traditional spectrogram-based methods, these operate directly on raw waveform inputs, enabling end-to-end optimization and low-latency processing and preserving phase information. 

\begin{figure}[tbp!]
    \centering
    \includegraphics[width=0.9\linewidth]{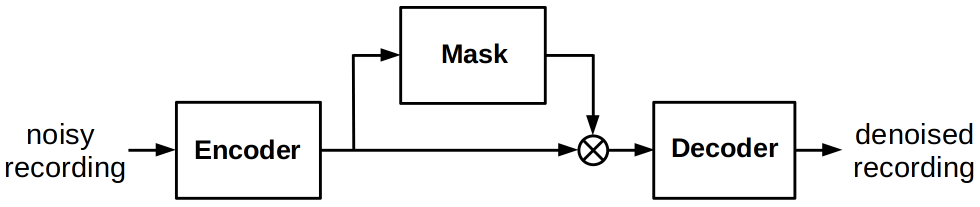}
    \caption{Conv-TasNet block diagram for soundscape denoising.}
    \label{fig:convtasnet}
\end{figure}

Conv-TasNet is a convolutional time-domain audio separation network originally developed for speech separation. It uses a learned encoder to map the input waveform into a latent representation, a temporal convolutional network to estimate a multiplicative mask over this representation, and a learned decoder to synthesize the denoised waveform. Conv-TasNet's ability to model long-range temporal dependencies makes it highly effective for single-channel speech enhancement tasks. We create a representation of a time-series recording contaminated with noise by employing a linear encoder. The separation process involves the multiplication of a trainable mask with the encoded input. The resultant masked encoded input is then transformed back into a denoised time-series recording via a linear decoder. This processing pipeline is illustrated in Fig.~\ref{fig:convtasnet}. WaveUNet is a waveform-domain U-Net originally proposed for music source separation; it learns multi-scale temporal features through successive downsampling and convolution, and reconstructs the output via a symmetric upsampling path with skip connections that preserve fine time-scale structure while suppressing noise. DEMUCSv2 is a waveform-domain encoder--decoder with strided convolutions and skip connections, augmented with a sequence-modeling bottleneck (e.g., recurrent layers) to better capture long-range temporal context for reconstruction. While these architectures were initially developed and validated on speech and music, their core design choices—1D time-domain processing, multi-scale feature extraction, and explicit modeling of temporal context—are well suited to underwater acoustics, where noise and target signals often overlap in time and frequency and phase can be informative.

The training and architectures of the denoisers, data augmentation, and hyperparameter selection are detailed in Appendix~A for ease of reading. Based on the comparison between the three architectures, Conv-TasNet is found to perform consistently better, and hence we choose this architecture for the denoiser for application in the rest of this work.

\begin{figure}[tbp!]
    \centering
    \includegraphics[width=1.0\linewidth]{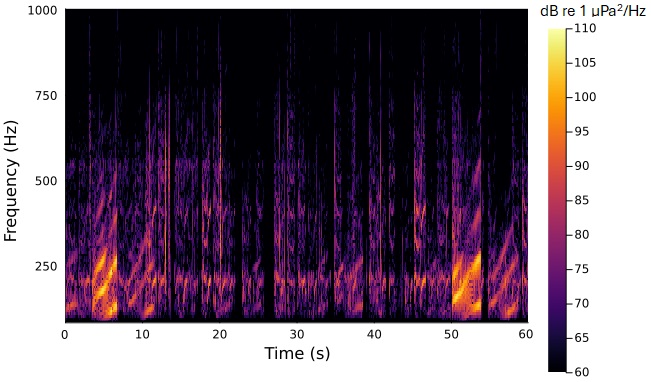}
    \caption{Spectrogram illustrating the denoised version of the acoustic data shown in Fig.~\ref{fig:spectrograms}(b), capturing a vibrant fish chorus.}
    \label{fig:spectrograms:convtasnet_fishchorus}
\end{figure}

\begin{figure}[tbp!]
\centering
\subfloat[Noisy recordings.\label{fig:roc:noisy}]{%
    \includegraphics[width=0.4\textwidth]{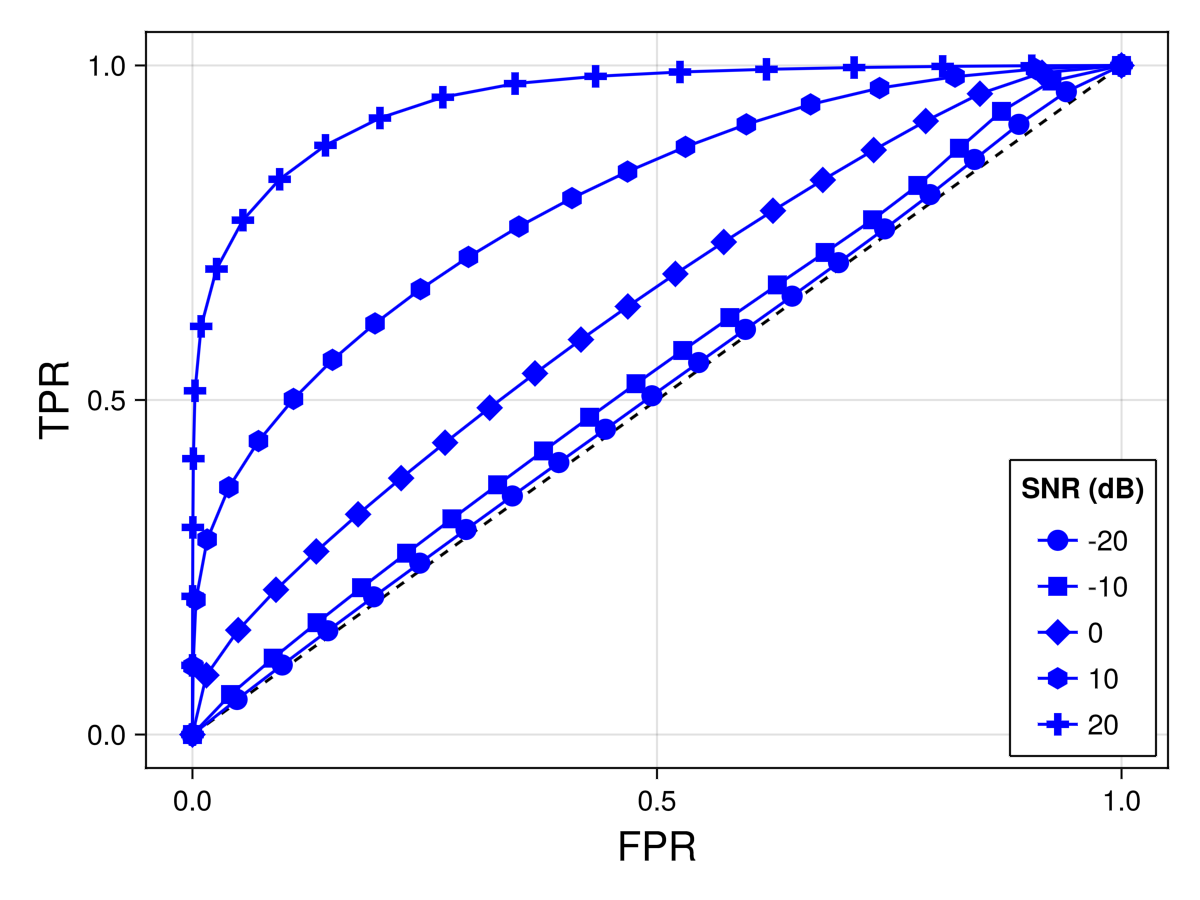}%
}%
\\
\subfloat[Denoised recordings.\label{fig:roc:denoised}]{%
    \includegraphics[width=0.4\textwidth]{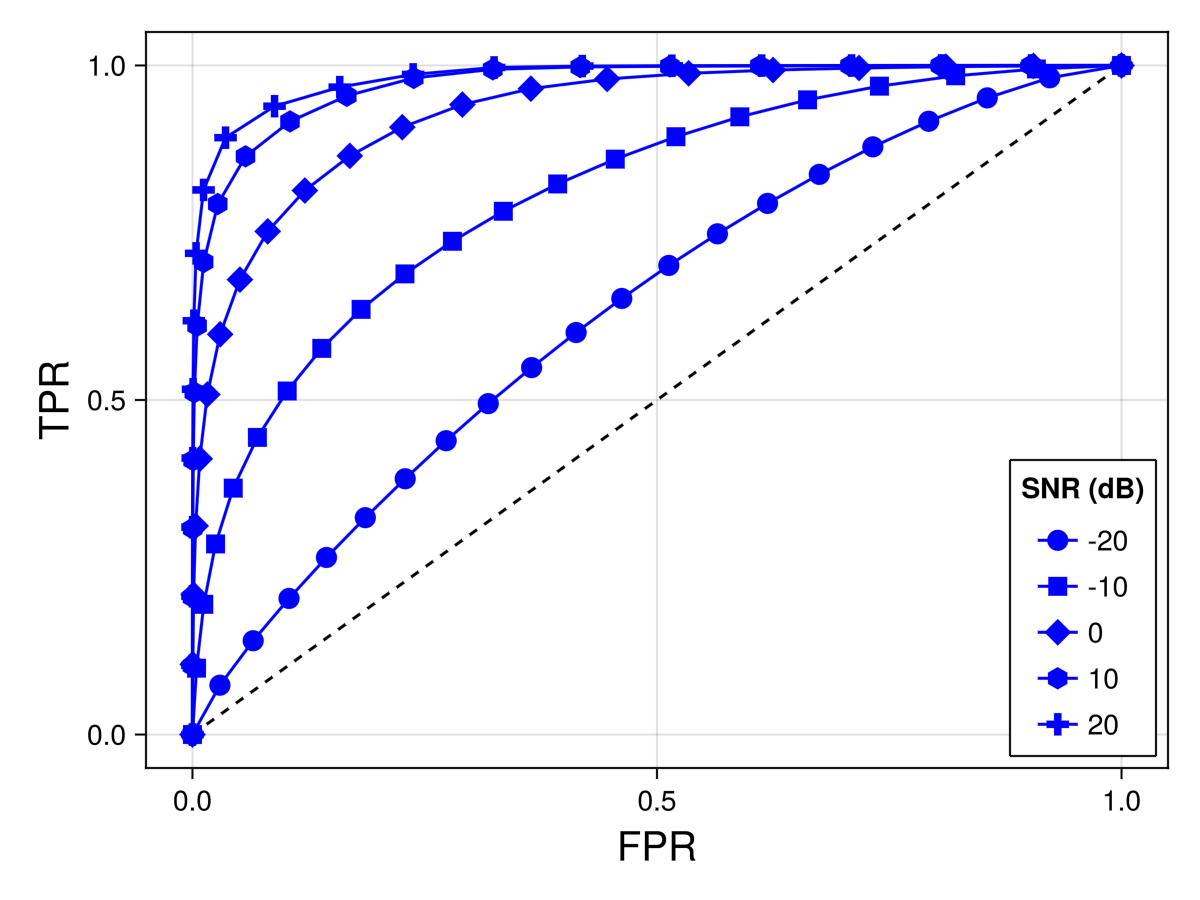}%
}
\caption{Quantifying denoiser performance in terms of ROC curves for detecting biological sounds of interest within the data, (a) without using and (b) with use of, the Reef denoiser. ROC curves that are more convex (higher than and away from the TPR = FPR dashed line) with larger area under the curve show better detector performance. The curves in (b) are seen to be consistently better than the corresponding ones in (a), showcasing the improvement due to the denoiser.}
\label{fig:roc}
\end{figure}
\subsection{Assessing denoiser performance}
We first qualitatively showcase the performance of the Conv-TasNet denoiser by highlighting three examples, in the form of spectrograms of the denoised outputs in Fig.~\ref{fig:spectrograms:convtasnet_fishchorus},~\ref{fig:spectrograms:shipnoise}(b), and \ref{fig:spectrograms:flownoise}(b). A comparison of these to the corresponding noisy spectrograms in Figs.~\ref{fig:spectrograms}(b), \ref{fig:spectrograms:shipnoise}(a), and \ref{fig:spectrograms:flownoise}(a), show that the denoiser is able to suppress the noise and retain just the fish vocalizations in these clips. In instances of low signal-to-noise (SNR) recordings, the denoiser resorts to a blanking process on the data. For the fish chorus recording in Fig.~\ref{fig:spectrograms:convtasnet_fishchorus}, the denoiser demonstrates efficacy in preserving fish vocalizations with minimal alteration. In \ref{fig:spectrograms:shipnoise}(b), the denoiser effectively suppresses the ship noise, unveiling a fish vocalization at around 25~s in the denoised recording. Similarly, the denoiser completely eliminates the current-induced noise in Fig.~\ref{fig:spectrograms:flownoise}(b), exposing four fish vocalizations in the denoised recording.

We now quantitatively assess the performance of the denoiser in terms of how it improves the detectability of signals of interest (fish vocalizations) embedded in the data. This is evaluated via the receiver operating characteristics (ROC), following the Neyman-Pearson criterion. The ROC is a plot of the variation of the true positive rate (TPR), or recall, versus the false positive rate (FPR), at each SNR value. Details of how these are computed are given in Appendix~B. Fig.~\ref{fig:roc} presents the ROC curves showcasing the detection performance on the (a) noisy and (b) denoised recordings across a range of SNRs. ROC curves that are more convex with a larger area under the curve generally indicate better detection performance. The detectability of the signals of interest are seen to improve with SNR, which is on expected lines. In all instances, the denoised recordings are seen to be consistently more detectable compared to the noisy recordings - this is visible in the form of the curves in Fig.~\ref{fig:roc}(b) being more convex compared to the corresponding ones in (a), and having more area under the curve. This improvement highlights the effectiveness of the denoising process in enhancing the detectability of biological sounds. 

\subsection{Results}
Having established the efficacy of the Reef denoiser, it is employed on the recorded data in the low-frequency band to obtain denoised recordings. Fig.~\ref{fig:spl}(c) illustrates the average low-frequency SPL computed from the denoised recordings over a 24-hour period. It is noteworthy that the distinctive double-peak pattern denoting the morning and evening choruses, is now revealed at the majority of coral reef sites, in comparison to \ref{fig:spl}(b). This is attributed to the effective denoising by the Reef denoiser that has removed a considerable portion of the shipping and flow-related noise. Notably, the acoustic chorus is observed to be louder during dusk as compared to dawn hours, and the activity is generally heightened at night compared to daytime by about two orders of magnitude, similar to that observed in \cite{Lin2023}. 


To further showcase the effect of the denoiser qualitatively, we now compare the average low-frequency SPL across dates and times based on the recordings before and after denoising in Fig.~\ref{fig:datehour}. After denoising, the data reveals the presence of morning and evening choruses in the form of elevated sound levels between 6 PM to 6 AM. The patches attributed to current-induced noise, as observed in Fig.~\ref{fig:datehour}(a) and (b), are largely mitigated in (c) and (d), with heightened SPLs recorded during the morning and evening hours. 
\section{Assessment in terms of acoustic indices}  \label{section:pam}
In this section, we examine the relationship between the acoustic data and coral reef health parameters, evaluating the usefulness of acoustic indices in reef health assessment and the potential for using these indices in long-term acoustic monitoring. 

\begin{figure}[tbp!]
    \centering
    \includegraphics[width=0.45\textwidth]{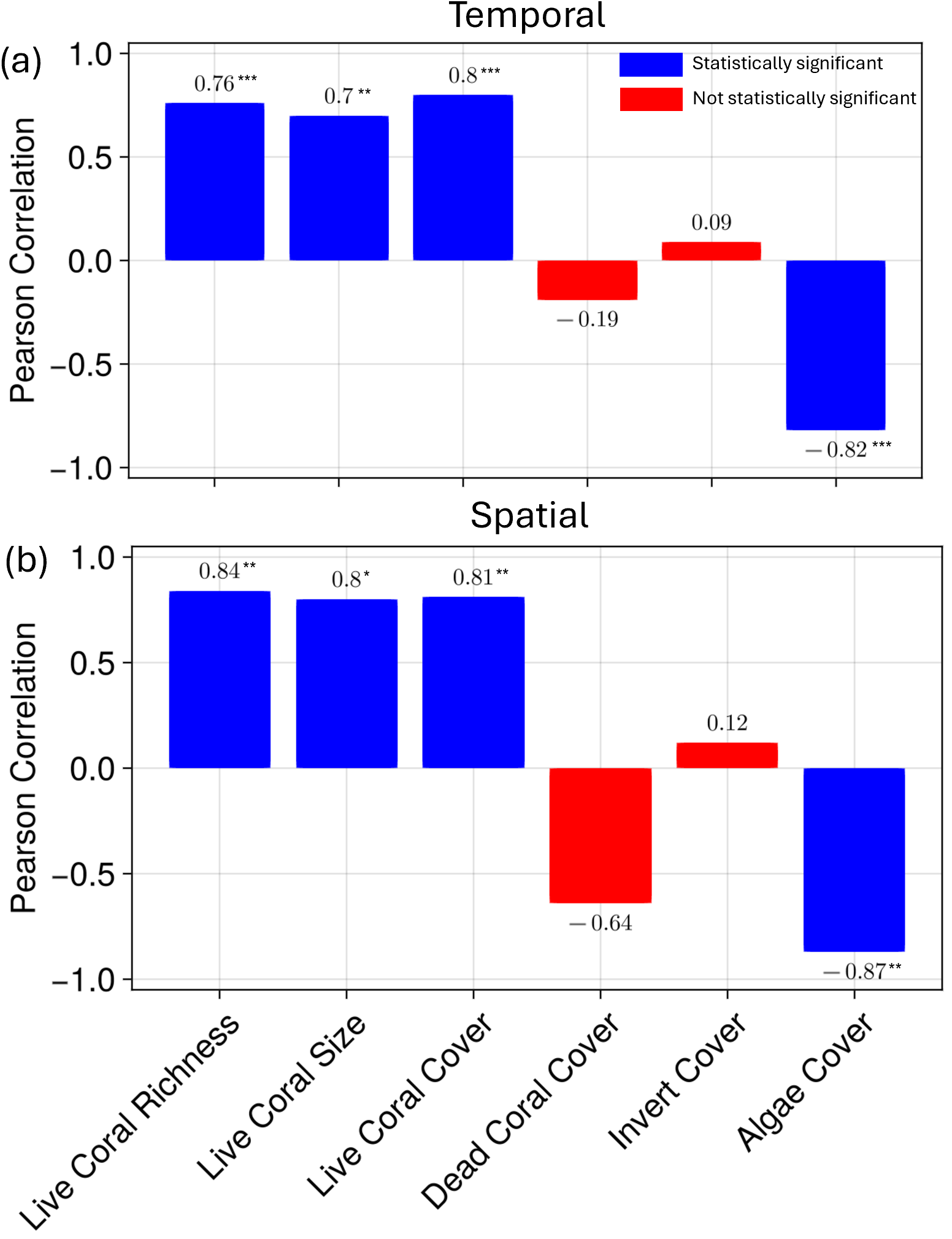}
\caption{(a) Temporal and (b) spatial correlation between reef parameters and snap rate in the high-frequency band.}
\label{fig:spl_transect_variables_highfreq}
\end{figure}

\subsection{Computation of indices}
The acoustic indices analyzed include the snapping shrimp snap rate, SPL (computed separately in the low-frequency and high-frequency bands) and Acoustic Complexity Index (ACI), of which the latter two have been defined and established in the bioacoustics literature. The snap rate quantifies the shrimp activity and is a particularly useful metric for the snapping shrimp-dense waters of Singapore. A similar index has been explored in \cite{nedelec2015soundscapes}. This index is estimated by first detecting snaps - this involves computing the envelope of the acoustic pressure signal via a Hilbert transform, and then thresholding it to retain values only within the top 0.1$^\text{th}$ percentile, based on \cite{Chitre2012}. The intuition is that these samples are expected to correspond to strong snaps. All peaks exceeding this threshold are identified as strong snaps, and the number of snaps are divided by the time duration of the clip (in minutes) to obtain the snap rate in snaps per minute.

SPL (computed in dB rms re 1$\mu$Pa) reflects the overall intensity of sound, and the ACI gauges the richness and variety of sounds within the soundscapes~\cite{pieretti2011new}. Previous studies have demonstrated positive correlations between SPL and live coral cover~\cite{bertucci2016acoustic}, and between ACI and fish diversity~\cite{harris2016ecoacoustic,bertucci2016acoustic}, underscoring their utility in capturing the intricacies of reef ecosystems~\cite{harris2016ecoacoustic,bertucci2016acoustic}. The SPL is defined (in dB re 1$\mu$Pa) as
\begin{equation}
    \text{SPL} = 20 \log_{10}\left(\frac{E_\text{rms}}{E_\text{ref}}\right)
\end{equation}
where $E_\text{ref}$ is the reference pressure of $10^{-6}$Pa, and
\begin{equation}
    E_\text{rms} = \sqrt{\frac{\sum_{j=1}^{m} x^2(j)}{m}}
\end{equation}
where $m$ is the number of samples in the recording, and $x(j)$ is the $j^{th}$ sample of the recording timeseries in units of Pa.

The ACI, a dimensionless index, is computed based on spectrograms of the sound data, segmented temporally. The index is defined as~\cite{pieretti2011new}
\begin{equation}
    \text{ACI} = \sum_{l=1}^{q} \sum_{j=1}^{m} \frac{\sum_{k=1}^{n} \left| I_k - I_{k-1} \right|}{\sum_{k=1}^{n} I_k},
\end{equation}
where $k$, $j$ and $l$ are the temporal step, temporal segment and frequency bin indices respectively; $I_k$ is the intensity in the $k^\text{th}$ step; $n$ is the number of temporal steps within each temporal segment considered for the spectrogram; $m$ is the number of temporal segments considered in the entire recording; and $q$ is the total number of frequency bins considered in the computation. For the ACI, a 0.128~s time-step and 50\% overlap between steps with a Hanning window were used for spectrogram computation, resulting in a frequency resolution of 7.8~Hz. Daily averages of the indices were computed by averaging the values computed for 1-minute segments. 

\FloatBarrier

\begin{figure}[tbp!]
    \includegraphics[width=.51\textwidth]{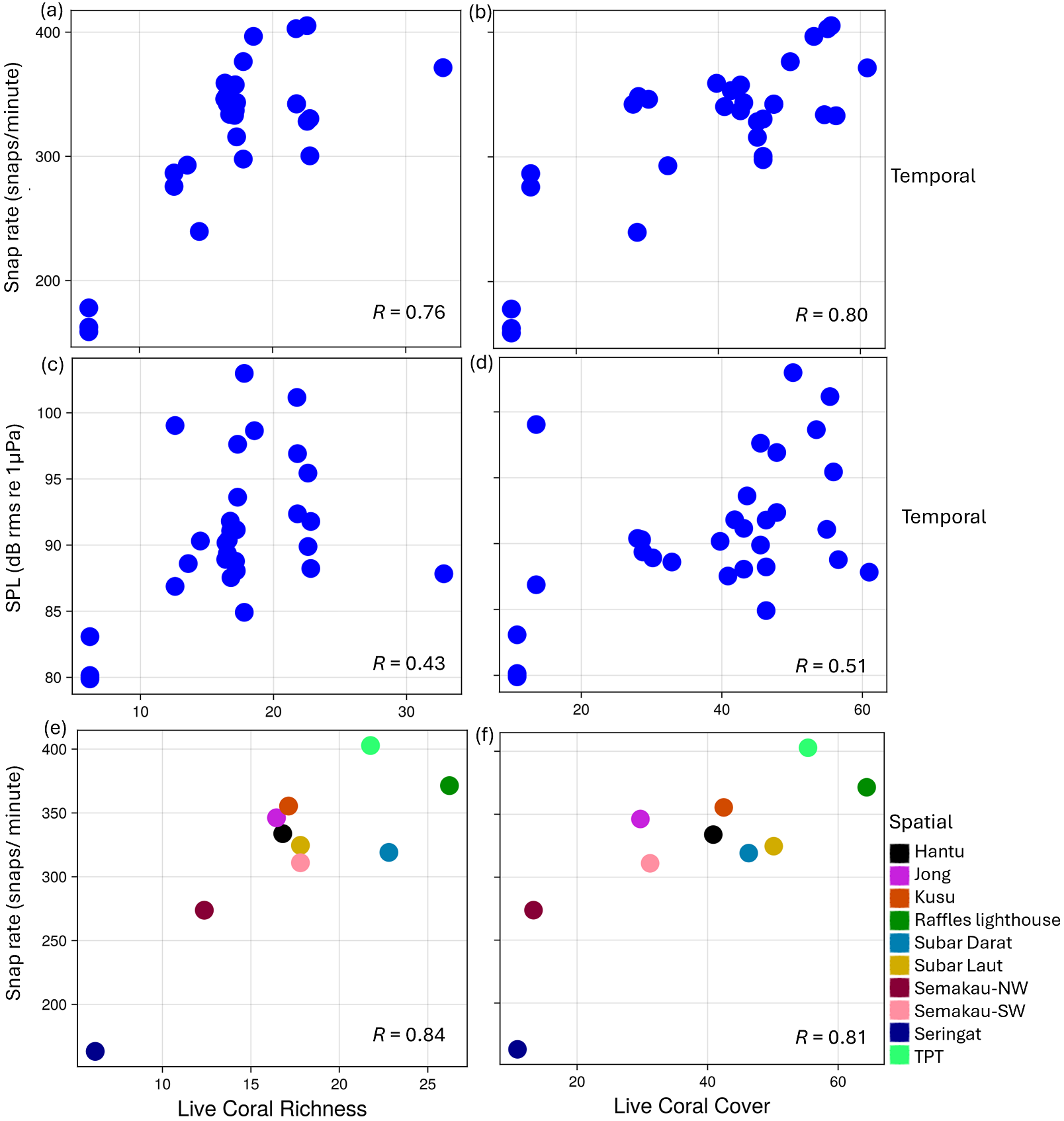}
\caption{Scatter plot of (a), (b), (e) and (f) snap rate, and (c) and (d) low-frequency SPL, against live coral richness and live coral cover. (a)-(d) show the temporal correlation, whereas (e) and (f) capture the spatial correlation.}
\label{fig:correlation}
\end{figure}
\begin{figure}[tbp!]
    \centering
    \includegraphics[width=0.45\textwidth]{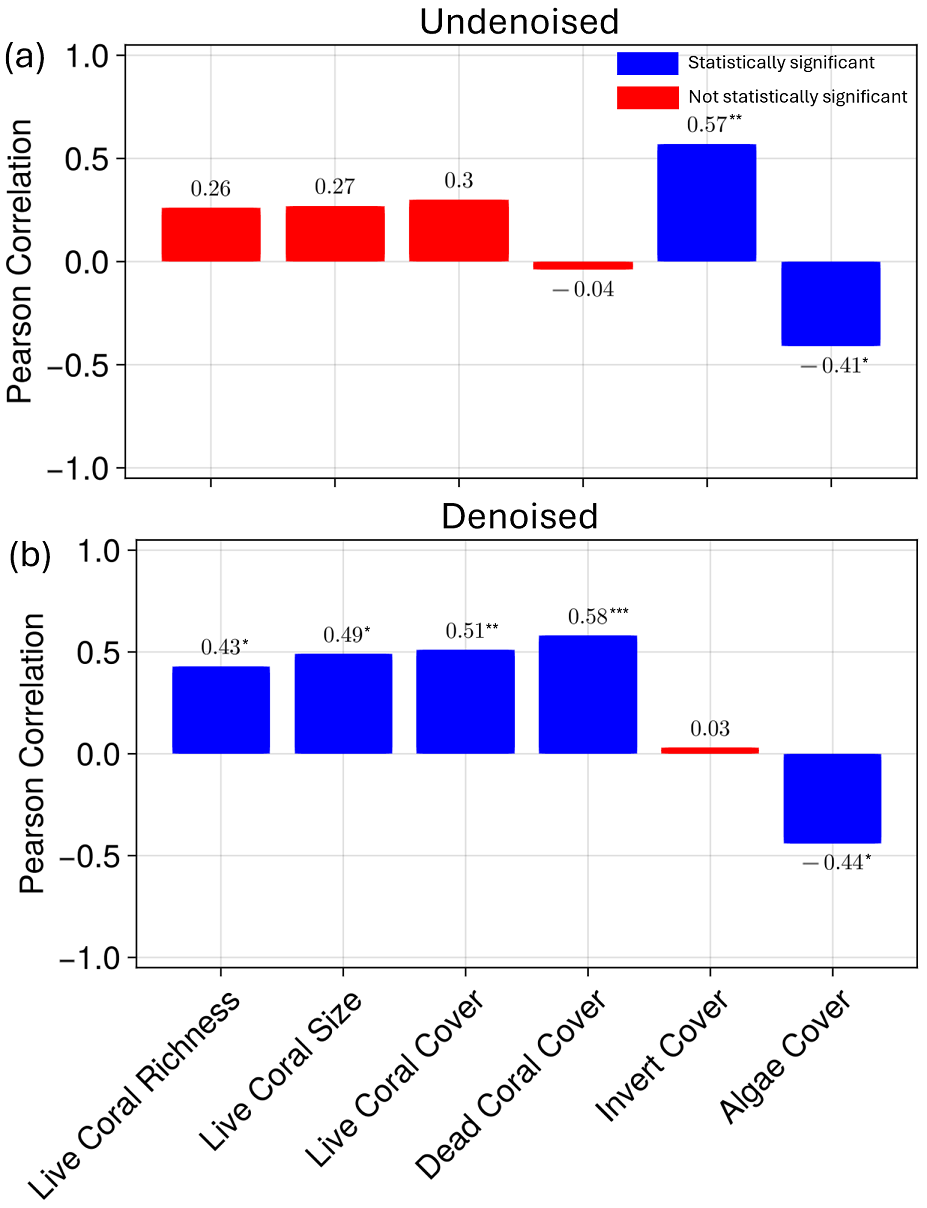}
\caption{Temporal correlation between transect variables and SPL in the low-frequency band with (a) noisy and (b) denoised acoustic data.}
\label{fig:spl_transect_variables}
\end{figure}
\subsection{Correlation of acoustic indices and reef health parameters}\label{sec:correlationmethods}
To evaluate the utility of acoustic indices in assessing reef health, we now analyze the correlation of these indices against the reef parameters measured via transect surveys. One challenge here is that the acoustic data and transect data were not always overlapping in time, because the datasets were collected intermittently when deployment logistics allowed. However, based on insights from coral reef biology, the reef health parameters are not expected to fluctuate significantly within the sampling periods, which allows us to interpolate this data linearly within these periods in order to facilitate a comparison against the acoustic data. While this does not necessarily capture any fine-scale variations, this gives us the best possible idea of how well-correlated the acoustics is to the reef parameters based on the data collected.

From the low-frequency band, we consider the SPL and ACI indices, and compare the effects on these indices before and after denoising. Henceforth, the SPL refers to the low-frequency band, unless otherwise mentioned. From the high-frequency band, we correlate mostly the snap rate, making observations with the high-frequency SPL in a few cases, and no denoising is applied. We use the Pearson correlation coefficient to quantify the relationships between the acoustic indices and reef parameters, with statistical significance assessed using a t-test. Correlations with $p$-values below $0.05$ are considered significant. The notations $^*$, $^{**}$ and $^{***}$ denote values with $p<0.05$, $p<0.01$ and $p<0.001$ respectively.

\subsection{Results and Discussion}
\subsubsection{Overall temporal correlation}
We first examine the correlation between the indices and reef parameters at different times averaged over all the sites, to examine whether the indices can capture temporal variability in reef health. Fig.~\ref{fig:spl_transect_variables_highfreq} plots the correlation coefficients between the reef parameters and the shrimp snap rate in the high-frequency band (1-20~kHz). Statistically significant correlations are denoted in blue, and insignificant ones in red. The snap rate shows strong positive correlations with live coral richness (correlation coefficient $R$=0.76), size ($R$=0.7), and cover ($R$=0.8). The correlation with the coral richness and cover are also showcased in the scatter plot in Fig.~\ref{fig:correlation}~(a) and (b). 

Most of these correlations are significant, with the exception of invertebrate and dead coral cover. This establishes that snap rate reliably tracks temporal variation in live coral health within a similar geographic region. Reefs with higher live coral cover tend to support greater biodiversity~\cite{coker2012interactive}, reflected in elevated snapping shrimp activity. The snap rate is also negatively correlated with the total algal cover ($R$=-0.82). This is likely because higher algal cover in a region denotes poor reef health as a result of algae taking over sites previously covered by reefs, or due to competitive exclusion by algae over reefs, thus indicating reduced biodiversity and biological activity, and consequently fewer sound-producing organisms~\cite{Fong2021, Fong2023}. High-frequency SPL is moderately correlated with the live coral size ($R$=0.58, statistically significant), but less so with other parameters.

\begin{figure}[tbp!]
    \includegraphics[width=.45\textwidth]{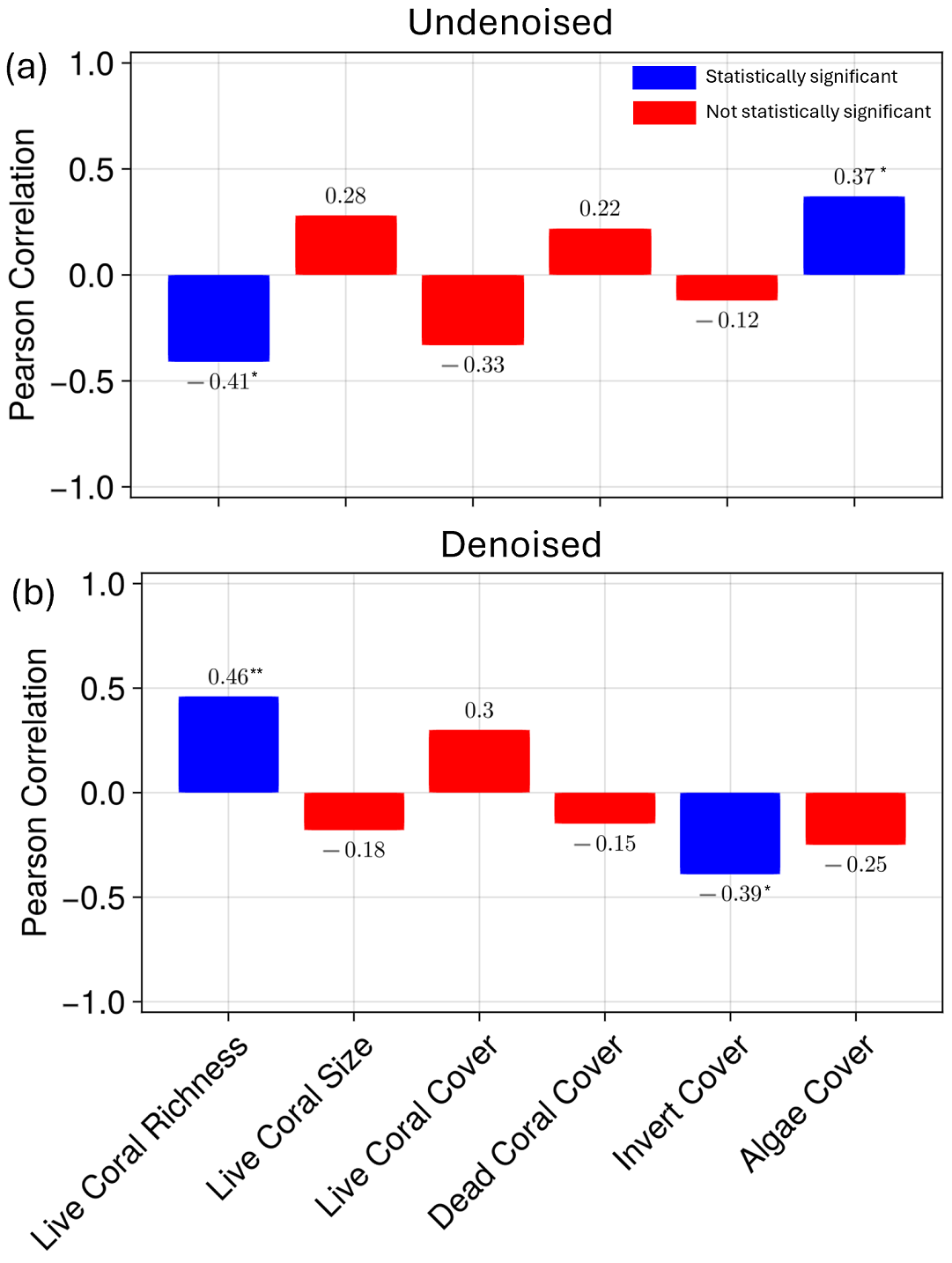}
\caption{Temporal correlation between transect variables and ACI in the low-frequency band with (a) noisy and (b) denoised acoustic data.}
\label{fig:aci_transect_variables}
\end{figure}

Fig.~\ref{fig:spl_transect_variables} plots the correlations between the transect variables and low-frequency SPL, (a) before and (b) after denoising. The SPL of the noisy data does not show any noteworthy correlation with reef parameters except with invertebrate cover and algal cover. In contrast, the denoised data exhibits a moderate correlation with the live coral richness ($R$=0.43), size ($R$=0.49), and cover ($R$=0.51), and dead coral cover ($R$=0.58), all statistically significant. Correlations with the live coral richness and coral cover are also shown in Fig.~\ref{fig:correlation}~(c) and (d), and indicate generally higher biodiversity being supported by coral reef cover. The correlation with dead coral cover could indicate the presence of grazers inhabiting such areas \cite{Korzen2011}. This index is also negatively correlated with the total algal cover ($R$=-0.44), as observed with the SPL in the high frequency band. 

The low-frequency SPL and ACI are indicative of bioacoustic contributions from vocalizing fishes inhabiting the reef region, which are associated with healthier reefs. However, these indices generally show lower correlation with coral parameters than the snap rate. This is likely because of two interconnected reasons, (1) masking by stronger contributions from anthropogenic noise in the lower frequency band, and (2) the louder, clearer contribution from impulsive shrimp snaps that are easier to pick up, making it a cleaner and robust indicator of bioacoustic activity. These observations match recent findings by Raick et al \cite{Raick2025}, who advocate for wider observation of high-frequency bioacoustic indicators from invertebrate activity.

\begin{figure}[tbp!]
\centering
    \includegraphics[width=.4\textwidth]{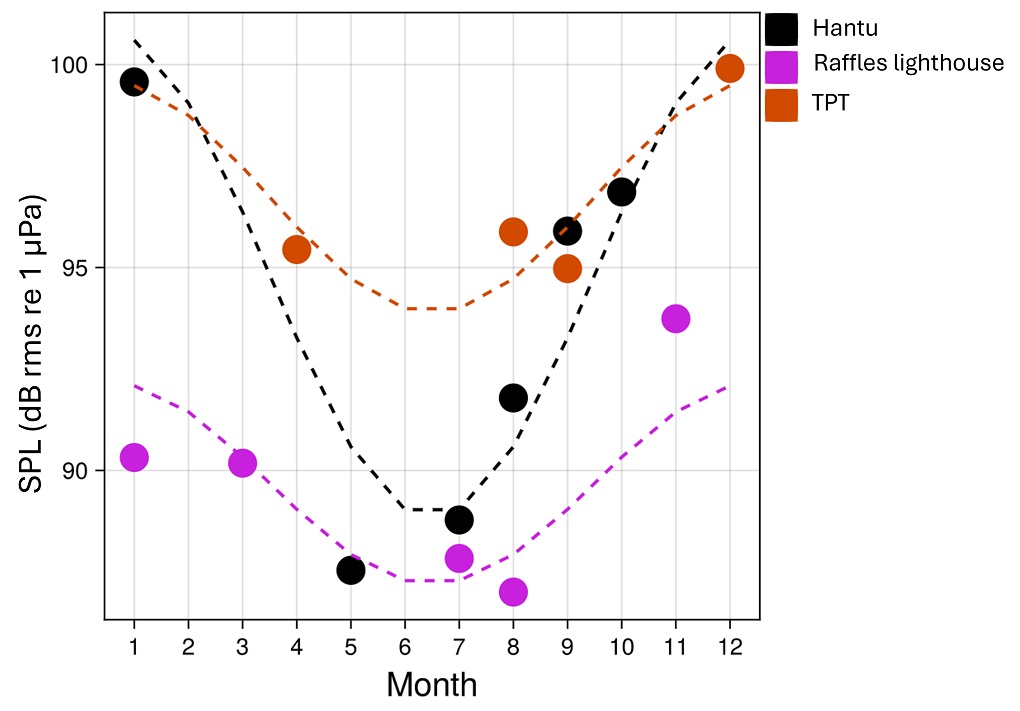}
\caption{Monthly-averaged variation of low-frequency band SPL with denoised data at three sites, and best-fit annual cyclic trend for each of these sites.}
\label{fig:monthly_variation}
\end{figure}
Notably, denoising improves the magnitude of correlations between the low-frequency SPL and the live coral parameters, dead coral cover and algal cover. This underscores the effectiveness of the denoising approach in improving the usability of acoustic metrics, and highlights its usefulness in aiding acoustic assessment of temporal reef health changes. 

Fig.~\ref{fig:aci_transect_variables}~(a), shows that undenoised ACI in the low-frequency band displays counterintuitive negative temporal correlations with live coral richness ($R$=-0.41), live coral cover ($R$=-0.33), and a positive correlation with algal cover ($R$=0.37). These suggest that as live coral richness and cover increase, the ACI decreases, which contradicts typical expectations of a positive relationship between biodiversity and acoustic complexity. This is likely due to the strong non-biological noise in the raw data, which can mask biological sounds and skew the overall acoustic complexity measurements. Fig.~\ref{fig:aci_transect_variables}~(b) presents the correlation of the ACI with the denoised acoustic recordings. The positive correlations between ACI and live coral richness ($R$=0.46, statistically significant) and cover ($R$=0.3, not significant), are more consistent with expectations showing that the denoising has slightly improved the effectiveness of the ACI in the acoustic assessment. The ACI is most correlated with the coral richness, but not significantly correlated with the other parameters indicating reef coverage or volume. This is understandable because the ACI is an indicator of diversity rather than volume, and thus ideally placed to predict the variation in richness.

The improvement by denoising is more pronounced for SPL than ACI. This is because SPL is a simpler metric of the overall energy, which the denoiser cleans up effectively. The SPL is thus able to robustly represent bio-acoustic activity within the denoised recordings. The ACI, on the other hand, extracts finer spectral details from the fluctuations within each frequency band. While the denoising may improve the spectrogram quality to some degree, it may also distort some of these details or create minor artifacts that interfere with the ACI. Thus, the improvement in ACI due to denoising is not as significant as SPL. 

\subsubsection{Temporal variation at each site}
Having examined temporal variability averaged across sites so far, we now assess temporal correlations of acoustics with reef parameters at individual sites. As mentioned earlier, within the timespan of this study, we do not have enough samples at each site to understand the finer temporal fluctuations in most of the reef parameters satisfactorily. However, we have collected more samples for the macroalgae parameter. Furthermore, for this parameter, a cyclical variation over the year is expected due to seasonal variation in water temperature, light, etc, with seawater temperature considered the most important for driving \textit{Sargassum} seasonal growth and reproductive cycles~\cite{Low2019}. This would have an annual maximum expected around December-January, and a minimum around June-July~\cite{Low2019}. Based on this intuition, for the macroalgal coverage parameter, we fit a cyclical variation of the percentage cover $C$ over the year at each site, defined as
\begin{equation}
    C(d) = A \cos\left(\frac{2\pi (d - \phi)}{365} \right) + B
\end{equation}
where $d$ is the day of the year, $A$, $B$ and $\phi$ are site-specific parameters to be fit to the transect data. For $\phi = 0$, this function reflects a cyclical variation within $B\pm A$ with a peak occurring on July 2. This function is used to interpolate the data for the macroalgae parameter for each site separately, and correlated against the acoustic data. Given that we know the nature of fluctuations expected in this parameter, we can more conclusively test its correlation against the acoustics. 

Fig.~\ref{fig:monthly_variation} shows the monthly trend of denoised low-frequency acoustic data averaged over the two years at Hantu, Raffles Lighthouse and TPT. At Hantu, where the ambient noise level was minimal (Fig.~\ref{fig:spl}), the cyclic annual variation is especially clear. The cyclic pattern of macroalgal cover is strongly negatively correlated with low-frequency SPL at Hantu ($R=-0.92$, significant), and moderately at Raffles Lighthouse ($R=-0.84$) and TPT ($R=-0.83$), both not statistically significant. This negative correlation suggests that periods of high macroalgal cover coincide with lower bioacoustic activity, likely due to competitive exclusion or habitat alteration that reduces soniferous fauna~\cite{Fong2021, Fong2023}. This analysis also highlights the capability of acoustic indices to track and monitor temporal variation in environmental health.
\begin{table*}
\renewcommand{\arraystretch}{1.3}
    \caption{Coefficients for the composite acoustic index for monitoring temporal variations for each reef parameter. Statistically insignificant ($p>0.05$) values are marked in red.}
    \label{tab:multilineartemporal}
    \centering
    \begin{tabular}{|c|c|c|c|c|c|c|}
    \hline
        \textbf{Reef parameter} & \textbf{Snap-rate} - $a_i$ & \textbf{Low-frequency SPL} - $b_i$ & \textbf{ACI} - $c_i$ & \textbf{Intercept} - $d_i$ & $\mathbf{R}$ & \textbf{p-value}\\ \hline
        Live coral richness & 0.069$^{***}$ & \textcolor{red}{-0.104} & 0.03$^{**}$ & \textcolor{red}{-22.08} & 0.845$^{***}$ & $<10^{-4}$\\ \hline
        Live coral size & 0.117 $^{***}$& \textcolor{red}{-0.044} & -0.053$^{*}$ & \textcolor{red}{56.27} & 0.774$^{***}$ & $<10^{-4}$ \\ \hline
        Live coral cover & 0.200$^{***}$ & \textcolor{red}{-0.066} & \textcolor{red}{0.038} & \textcolor{red}{-53.42} & 0.815$^{***}$ & $<10^{-4}$\\ \hline
        Dead coral cover & \textcolor{red}{-0.064} & \textcolor{red}{1.46} & \textcolor{red}{-0.010} & \textcolor{red}{-80.8} & \textcolor{red}{0.352} & 0.206 \\\hline
        Invertebrate cover & 0.023$^{*}$ & -0.262$^{*}$ & -0.041$^{***}$ & 59.15$^{***}$ & 0.507$^{**}$ & 0.004\\ \hline
        Total algal cover & -0.169$^{***}$ & \textcolor{red}{0.343} & \textcolor{red}{-0.010} & \textcolor{red}{64.29} & 0.851$^{***}$ & $< 10^{-4}$\\ \hline
    \end{tabular}
\end{table*}
\begin{table*}
    \caption{Coefficients for the composite acoustic index for monitoring spatial variations for each reef parameter. Statistically insignificant ($p>0.05$) values are marked in red.}
    \label{tab:multilinearspatial}
    \centering
    \begin{tabular}{|c|c|c|c|c|c|c|}
    \hline
        \textbf{Reef parameter} & \textbf{Snap-rate} - $a_i$ & \textbf{Low-frequency SPL} - $b_i$ & \textbf{ACI} - $c_i$ & \textbf{Intercept} - $d_i$ & $\mathbf{R}$ & \textbf{p-value}\\ \hline
        Live coral richness & 0.061$^{*}$ & \textcolor{red}{0.034} & 0.035$^{*}$ & \textcolor{red}{-37.18} & 0.89$^{*}$ & $0.018$\\ \hline
        Live coral size & \textcolor{red}{0.100} & \textcolor{red}{0.291} & \textcolor{red}{-0.045} & \textcolor{red}{24.41} & \textcolor{red}{0.858} & $0.21$ \\ \hline
        Live coral cover & \textcolor{red}{0.209} & \textcolor{red}{-0.164} & \textcolor{red}{0.054} & \textcolor{red}{-62.19} & \textcolor{red}{0.83} & $0.058$\\ \hline
        Dead coral cover & \textcolor{red}{-0.122} & \textcolor{red}{0.788} & \textcolor{red}{-0.054} & \textcolor{red}{42.8} & \textcolor{red}{0.727} & 0.343\\\hline
        Invertebrate cover & \textcolor{red}{0.054} & \textcolor{red}{-0.643} & -0.079$^{*}$ & 120.1$^{*}$ & \textcolor{red}{0.786} & 0.103\\ \hline
        Total algal cover & -0.181$^{*}$ & \textcolor{red}{0.39} & \textcolor{red}{-0.020} & \textcolor{red}{72.5} & 0.883$^{*}$ & $0.021$\\ \hline
    \end{tabular}
\end{table*}

\subsubsection{Spatial correlation}
Finally, we examine the correlation between acoustics and reef parameters across  the ten different study sites, i.e., the spatial correlation. Fig.~\ref{fig:spl_transect_variables_highfreq}(b) shows the spatial correlation between snap-rate and reef parameters at different locations. The snap rate is significantly positively-correlated with the living parameters, namely the live coral richness ($R$=0.84), size ($R$=0.8) and cover ($R$=0.81), and significantly negatively correlated with algal cover ($R$=-0.87), following previous observations with the temporal correlation. The high-frequency SPL is weakly correlated with coral cover ($R$=0.54, not significant), and less correlated with other parameters. Low-frequency SPL exhibited weaker, non-significant spatial correlations, partly due the smaller sample size (10) for spatial analysis. Once again, this analysis showcases the usefulness of the snap rate as a useful feature to gauge reef health, this time, across different locations, and also its increased reliability over the low-frequency indices. Note that apart from differences in biological activity, the difference in the acoustic channel in which sound propagates can also influence sound levels, including factors such as the local bathymetry, sediment type and proximity to reflecting structures. These differences have not been incorporated into the indices in order to test correlation against biological activity. Despite this, the snap rate reflects the state of reef health across different sites within Singapore waters, showing its robustness.

\section{Composite acoustic index for reef health monitoring}
Having established the correlation of three acoustic indices with key reef parameters, we now propose their use as acoustic proxies for reef health monitoring. These indices—snap rate, denoised low-frequency SPL, and low-frequency ACI—are expected to capture complementary aspects of the reef soundscape, corresponding respectively to invertebrate (shrimp) activity, general bioacoustic activity (dominated by fishes), and acoustic diversity (reflecting species richness).

To quantitatively relate acoustic observations to ecological parameters, we construct composite acoustic indices $H_i$ as a linear combination of the three acoustic indices, with regression coefficients $a_{i}$, $b_{i}$, and $c_{i}$, and intercept $d_{i}$. For monitoring the temporal variations with time $t$, these composite indices are defined as
\begin{equation}
H_i(t) = a_i \text{Snap-rate}(t)+ b_i \text{SPL}(t) + c_i \text{ACI}(t) + d_i,
\end{equation}
whereas for monitoring spatial variations, the composite indices are defined as
\begin{equation}
H_i(\textbf{x}) = a_i \text{Snap-rate}(\textbf{x})+ b_i \text{SPL}(\textbf{x}) + c_i \text{ACI}(\textbf{x}) + d_i,
\end{equation}
where $\textbf{x}$ denotes location in Latitude-Longitude, $i\in \{1,2,3,4,5\}$ indicates each of the reef parameters of interest. The terms $a_i$, $b_i$, $c_i$, and $d_i$ are estimated via a multi-linear least-squares regression. We note that while the true relationships of the reef parameters with each of these indices could be non-linear, the linear approach taken here provides an interpretable and practical first-order model for proxy development.

For temporal variations, the coefficients for the composite indices for each parameter are given in Table~\ref{tab:multilineartemporal}, along with the regression coefficient for the composite index. The coefficients for spatial variation are given in Table~\ref{tab:multilinearspatial}. The results indicate that the composite indices have strong predictive power for temporal variation in live coral richness, size, cover, and total algal cover. There is a moderate, statistically significant correlation with invertebrate cover, whereas the composite index for dead coral cover does not show significant predictive power. The snap rate, in particular, consistently emerges as the strongest predictor across both temporal and spatial scales, re-emphasizing the diagnostic value of high-frequency snap activity as a proxy for live coral health and overall biodiversity. Its negative association with algal cover supports the ecological premise that healthier reefs with higher coral cover support more snapping shrimp, whereas degraded reefs with higher algal cover see reduced soniferous activity.

Low-frequency SPL, while moderately correlated with several of the reef parameters in isolation, adds limited additional predictive value beyond the snap rate in the composite models. This suggests some redundancy, but SPL remains informative for specific metrics such as invertebrate cover, likely reflecting complementary vocalizations from reef fishes. Furthermore, it is likely that the low-frequency SPL may help with prediction of additional aspects of reef health not considered here. 

The ACI does not contribute much to predicting the variability in most reef parameters (p-value of the ACI coefficient for coral size is borderline at $0.041)$, but it is most predictive for coral richness. This underscores that ACI is an indicator of diversity rather than volume, and thus more suitable to predict the variation in richness, though it is less correlated with and not helpful for predicting other reef parameters. This likely also indicates the increased acoustic diversity arising from the fish diversity as a result of increased coral richness. This shows the utility of this index in providing complementary information on the diversity in reef species.

For spatial variation (Table~\ref{tab:multilinearspatial}), similar patterns are observed as the temporal variation, with similar estimates of the coefficients for each of the acoustic indices also. The correlations are generally weaker or not statistically significant (with generally lower p-values also), particularly for coral size and cover, likely due to the limited spatial dataset. Nevertheless, strong spatial correlations are observed for composite indices of live coral richness and total algal cover.

\section{Conclusion}
Over a period of two years, we deployed monitoring equipment at 10 reef sites in the Singapore's waters, assembling an extensive archive of acoustic recordings encompassing a wide range of biological, physical, and anthropogenic sounds. Through analysis of this substantial dataset, we discovered that the coral reef soundscapes are consistently masked by noise from ships and tidal currents, especially in the lower frequency range, significantly masking and limiting the detectability of the biological sounds within the low-frequency soundscape. 

To address this issue, we trained a deep learning model (Conv-TasNet) to denoise the recordings. The results demonstrate that, once non-biological noises are suppressed using this Reef denoiser, the low-frequency reef soundscapes at many sites revealed distinct, vibrant biological patterns. The findings show that denoising can be a valuable tool for revealing biological sounds that may provide important insights into the state of coral reefs. By reducing the influence of non-biological noise, denoising helps to highlight the acoustic signals generated by the organisms inhabiting the reefs, thus improving its utility. This denoised acoustic data can facilitate more accurate monitoring in the trends of the state of coral reefs, and moving forward, assessments of the reef sites. This enhances the potential for reliable passive acoustic monitoring of coral reefs. 

We systematically evaluated three acoustic indices—snap rate, SPL, and ACI—for their value as proxies of reef health, benchmarking them against diver-based measurements of live coral richness, size, cover, invertebrate cover, and algal cover. Temporal correlations are recorded between the acoustic indices snap rate, sound pressure level, and acoustic complexity index, and the reef parameters namely live coral richness, size, and cover, invertebrate cover and algal cover. Snap rate consistently demonstrates the strongest and most robust correlation with coral parameters compared to the low-frequency indices, showcasing the increased effectiveness of high-frequency indicators (primarily from snapping shrimp) as sensitive indicators of live coral and overall reef biodiversity, aligning with Raick et al~\cite{Raick2025}. The snap rate also shows a strong spatial correlation with reef health, despite the change in environmental conditions across the sites, though the correlation of the low-frequency indices is not as strong. This further affirms the robustness of the high-frequency index used, namely, snap rate. So far, the SPL and ACI computed in the high frequency band have only yielded moderate or low correlations, which are in line with observations in the literature~\cite{Raick2025a} that these parameters may be sensitive to study site, depth or setting. A cyclic behavior is recorded in the low-frequency SPL variation at some sites, likely correlated with the macroalgal cover. Some relationships, such as negative correlations between ACI and invertebrate cover, point to the unrevealed complexity of the acoustic-ecological interface and warrant further investigation.

By combining these indices, we develop composite indices to explain the reef health variation with high fidelity using passive acoustics alone. These are strongly correlated to the reef health, both temporally and spatially. While some redundancy exists (e.g., between SPL and snap rate for some parameters), these indices may still provide different insights from complementary channels of information from the high and low frequency bands (arising from shrimp and fish vocalizations separately) for other aspects of reef health. The ACI is most effective in reflecting reef richness, showcasing its effectiveness in capturing the acoustic diversity which captures reef diversity.

In summary, while acoustic proxies may not fully replace traditional transect measures, our results show that they correlate well with, and can supplement, existing reef health assessments with additional channels of information. Thus, they have the potential to facilitate long-term, inexpensive, non-invasive and larger scale monitoring of reef health. In the light of recent disturbances in reef ecosystems in response to anthropogenic pressures, such large-scale solutions would be crucial to keep an eye (or ear) out on the state of the precious but fragile reef ecosystems. Advances in acoustic technologies such as distributed acoustic sensing may allow scaling up this monitoring by orders of magnitude and deliver synoptic, potentially basin‑level perspectives on reef health.

\appendices
\section{Training, data augmentation, architectures and hyperparameter selection for denoisers}
To train the denoisers through supervised learning, it is essential to have two distinct sets of recordings: the noisy recordings which, for our application scenario of interest, consist of a mixture of fish vocalizations, ship noise, and current-induced noise, and the corresponding clean recordings which exclusively contain fish vocalizations. Acquiring such noisy and clean recordings from the field can be quite challenging. However, obtaining recordings of each of these categories is feasible. Using this, we can synthesize noisy recordings by superimposing fish vocalizations, denoted as `signal data', with ship noise or current-induced noise, denoted as `noise data', in an additive manner. The composition of the signal and noise data available for training, validating and testing the denoiser is summarized in Table~\ref{table:datasets}. The signal data comprises $118$ fish sound snippets obtained from the FishSounds database~\cite{looby2023fishsounds}, varying in duration from $0.3$ to $300.0$ seconds. Additionally, from the acoustic recordings made by us in Singapore waters, identified as Reefwatch, we added $3663$ manually labeled fish sound snippets ranging from $1$ to $10$ seconds in duration. These sounds were annotated from a small portion of Reefwatch, marking common fish vocalizations such as foghorn, honk, drum, knock, and distress calls~\cite{looby2023fishsounds,lamont2022sound}. For the noise data, we used $569$ ship noise snippets with durations ranging from $6$ to $1887$ seconds, obtained from the DeepShip database~\cite{irfan2021deepship}, along with $218$ sound snippets from Reefwatch dominated by current-induced noise.

\begin{table}[tbp!]
\caption{Signal and noise datasets used for denoiser training.}
\label{table:datasets}
\centering
\begin{tabularx}{\linewidth}{|l|c|c|c|c|c|}
\hline
Type & Source & \multicolumn{3}{c|}{Number of recordings} & Clip durations (s) \\
\cline{3-5}
 & & Train & Validation & Test & \\ 
\hline\hline
\multirow{2}{*}{Signal} & FishSounds & $95$ & $23$ & $19$ & $0.3-300.0$ \\
 & Reefwatch & $3561$ & $102$ & $99$ & $1.0-10.0$ \\
\hline
\multirow{2}{*}{Noise} & DeepShip & $509$ & $60$ & $40$& $6.0-1887.0$ \\
 & Reefwatch & $202$ & $16$ & $10$ & $300.0$ \\
\hline
\end{tabularx}
\end{table}

Using these, we can now create pairs of noisy and clean recordings for developing the denoiser. To do this, $N$ signals are randomly chosen from the signal data and superimposed to generate a signal vector of a certain length $l$. $N$ was an integer selected randomly from the range $[1,5]$. Data augmentation, consisting of standard operations such as pitch shifting, time stretching, and tanh distortion was then applied to the signal vector. Each augmentation was performed stochastically with a probability of $0.5$. Simultaneously, we randomly select a noise recording from the noise data to construct a noise vector of length $l$. The result of these steps is a pair of signal and noise vectors, which are merged to produce a noisy recording, while the signal vector is retained as the clean recording. 

Due to memory constraints, we utilize a reduced-sized version of Conv-TasNet in this study. The mask includes two repeated 1D convolutional networks, with each network having $512$ channels in its convolutional layers. The detailed model structure is described in~\cite{luo2019conv}. The WaveUNet denoiser~\cite{Stoller2018} is implemented with a 12-level encoder/decoder structure, convolutional kernel size of 15 in the encoder and 5 in the decoder, and a linearly activated output head consisting of 1 channel output and base channel width of 48 with the number of channels growing by 48 in each layer. The DEMUCSv2 denoiser architecture~\cite{Defossez2021} employed by us consists of 6 encoder/decoder stages, with 64 base channels (width), kernel size of 8, stride of 4, two long short-term memory (LSTM) layers, and a resampling factor of 2 at both the input and output ends. The hyperparameters for these models were undertaken based on a brief ablation study.

For batched training, the denoiser processes sets of $32$ noisy recordings and generated denoised recordings. During training, the mean absolute error between the denoiser's timeseries output and the corresponding clean recording is minimized, using the Adam optimizer. The model is trained for $5000$ epochs, with $l$ chosen such that the segments are $10$-seconds long. The data is partitioned into training, validation, and test sets as outlined in Table~\ref{table:datasets}. Training is stopped when the validation score starts dropping, and the network with the best validation score is used for testing. It is important to note that the Reefwatch data used for training and validation constitutes only a tiny fraction of the entire set of acquired acoustic recordings. The remaining, unseen Reefwatch data is reserved for evaluation in the subsequent section.

The trained denoisers are compared in terms of two metrics: (1) output SNR (in dB) and (2) Scale-Invariant Signal-to-Distortion Ratio (SISDR). The output SNR of the denoiser output waveforms $\hat{\textbf{s}}$ are calculated against the known, undistorted signal $\textbf{s}$ as 
\begin{equation}
    \text{SNR$_{output}$} = \frac{|| \hat{\textbf{s}} ||^2}{||\textbf{s} - \hat{\textbf{s}} ||^2}
\end{equation}

\begin{figure}[tbp!]
\centering
    \includegraphics[width=.4\textwidth]{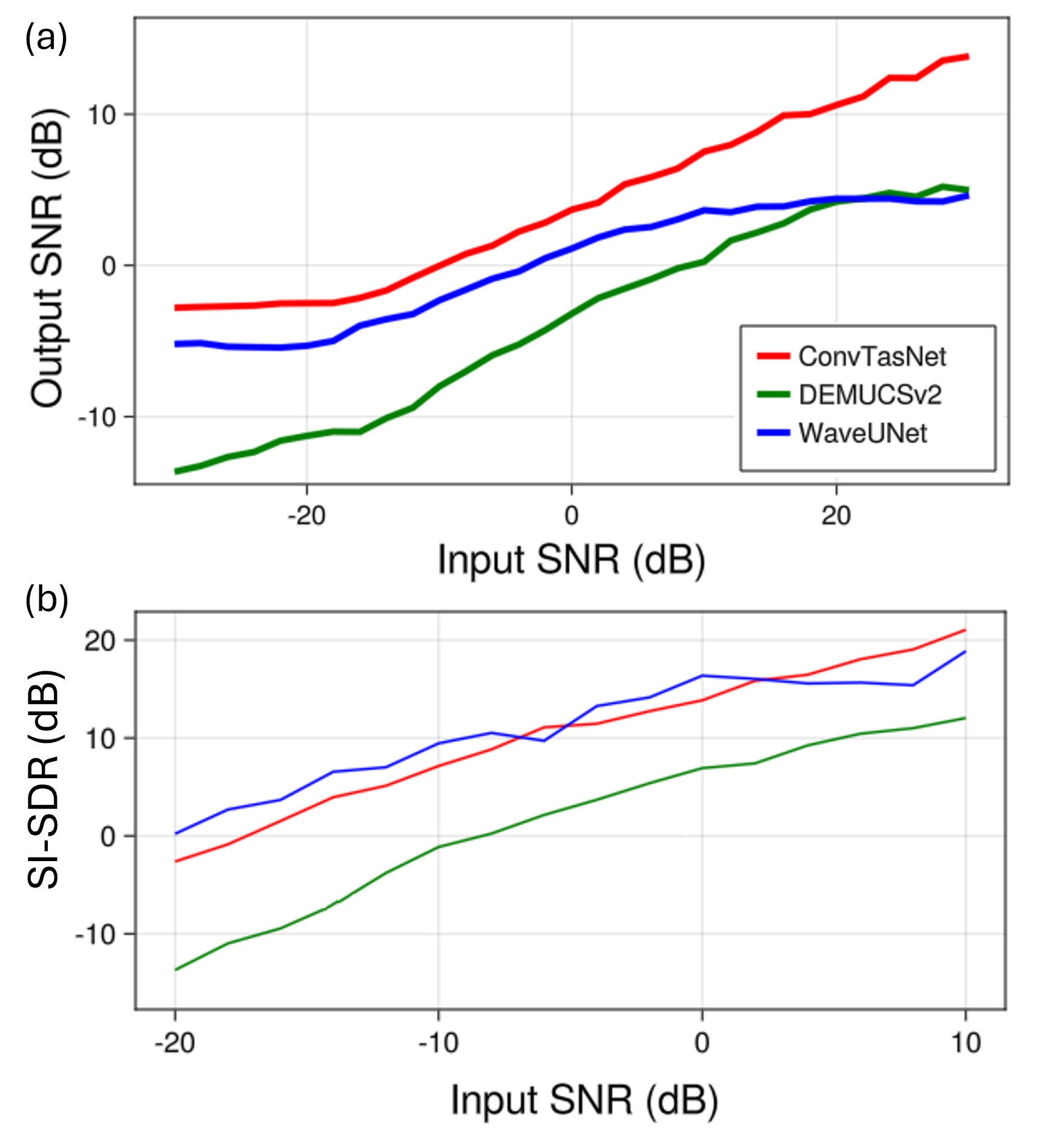}
\caption{Comparison of three denoiser configurations tested in terms of (a) Output SNR and (b) SISDR, against variation in input SNR (in dB).}
\label{fig:denoisercomparison}
\end{figure}

The SISDR quantifies the ratio of the energy of the scaled clean target signal to the energy of the residual noise, and is defined as 
\begin{equation}
    \text{SISDR$_{output}$} = \frac{|| a \textbf{s} ||^2}{||a \textbf{s} - \hat{\textbf{s}} ||^2}
\end{equation}
where $a = \frac{\textbf{s}^T\hat{\textbf{s}}}{|| \hat{\textbf{s}} ||^2}$. This metric measures the ratio of target signal energy to distortion energy while ignoring global amplitude scaling differences. Figure~\ref{fig:denoisercomparison} compares the performance of our three trained denoisers based on each of the architectures compared, in terms of output SNR and SISDR, at different values of SNR of the input waveform to the denoiser (in dB). The Conv-TasNet denoiser is shown to perform the best amongst the three in terms of output SNR, whereas in terms of SISDR, Conv-TasNet and WaveUNet perform equally well and better than DEMUCSv2. Hence, we select Conv-TasNet for application in this work.

\section{Quantifying denoiser performance in terms of ROC}
Denoiser performance is evaluated in terms of the detectability of the envelope of the signal within the data, normalized to lie within the range [0, 1]. The normalization is done for comparability across clips from different locations which can have differing sensor sensitivities and ambient noise backgrounds. In essence, this is equivalent to a detector decision rule that energy within a clip that exceed a certain detection threshold (as decided by the FPR) is a detection. Thus, the detectability of the signals are assessed in terms of the percentage of signal samples of interest within the data which exceed this threshold. 

To do this, we use the test sets from the Reefwatch data. We first compute the envelopes of the data with the clean, noisy, and denoised recordings from the test sets using the Hilbert transform. The envelope of each clip is then normalized to a range between $0$ and $1$. Within the clean recordings, signal envelope values above $0.01$ are classified as signal events to be detected. The samples within the normalized envelopes from the noisy and denoised recordings which cross a detector threshold are considered as detected events. The detected events which overlap with the signal events are counted as true positives, whereas those which do not overlap with signal events are counted as false positives. These are then compared to generate the ROC curves. 

The ROC curve is a plot of the TPR against the FPR, which helps assess detector performance as per the Neyman-Pearson criterion. The TPR and FPR are defined as
\begin{equation}
    \text{TPR} = \frac{\text{No. of True Positives}}{\text{No. of True Positives + No. of False Negatives}},
\end{equation}
\begin{equation}
    \text{FPR} = \frac{\text{No. of False Positives}}{\text{No. of False Positives + No. of True Negatives}}.
\end{equation}

\section*{Code Availability}
The neural network architecture and weights of the Conv-TasNet denoiser will be made available on reasonable request.

\section*{Acknowledgements}
This work was possible due to the support of grants from the National Research Foundation, Prime Minister's Office, Singapore under its Marine Science Research and Development Program MSRDP P-20. We thank the Reef Ecology Study Team for assisting with the coral reef visual survey.

\bibliographystyle{IEEEtran}
\bibliography{reference}

@techreport{burke2011reefs,
  title={Reefs at risk revisited},
  author={Burke, Lauretta and Reytar, Kathleen and Spalding, Mark and Perry, Allison},
  year={2011},
  institution ={World Resources Institute},  address   = {Washington, DC, USA},
  number = {ISBN: 978-1-56973-762-0},
  url = {https://www.wri.org/research/reefs-risk-revisited}
}

@incollection{chou2000southeast,
  title={{Southeast Asian reefs-status update: Cambodia, Indonesia, Malaysia, Philippines, Singapore, Thailand and Vietnam}},
  author={Chou, Loke Ming},
  booktitle={Status of coral reefs of the world: 2000},
  pages={117--129},
  year={2000},
  publisher = {Australian Institute of Marine Science},
  edition   = {First edition},
  editor = {C. Wilkinson},
  url       = {https://icriforum.org/wp-content/uploads/2019/12/gcrmn2000.pdf},
}

@article{Fong2023,
  author = {Fong, Jenny and Tang, Peggy P. Y. and Deignan, Lindsey K. and Seah, Jovena C. L. and McDougald, Diane and Rice, Scott A. and Todd, Peter A.},
  doi = {10.3390/microorganisms11092261},
  issn = {2076-2607},
  journal = {Microorganisms},
  month = {sep},
  number = {9},
  pages = {2261},
  title = {Chemically Mediated Interactions with Macroalgae Negatively Affect Coral Health but Induce Limited Changes in Coral Microbiomes},
  url = {https://doi.org/10.3390/microorganisms11092261},
  volume = {11},
  year = {2023}
}

@article{Fong2021,
  author = {Fong, Jenny and Todd, Peter A.},
  doi = {10.1016/j.marpolbul.2021.112849},
  issn = {0025326X},
  journal = {Marine Pollution Bulletin},
  month = {nov},
  pages = {112849},
  title = {{Spatio-temporal dynamics of coral–macroalgal interactions and their impacts on coral growth on urbanised reefs}},
  url = {https://doi.org/10.1016/j.marpolbul.2021.112849},
  volume = {172},
  year = {2021}
}

@article{Korzen2011,
  author = {Korzen, Leor and Israel, Alvaro and Abelson, Avigdor},
  doi = {10.1155/2011/960207},
  issn = {1687-9481},
  journal = {Journal of Marine Biology},
  pages = {1--8},
  title = {Grazing Effects of Fish versus Sea Urchins on Turf Algae and Coral Recruits: Possible Implications for Coral Reef Resilience and Restoration},
  url = {https://doi.org/10.1155/2011/960207},
  volume = {2011},
  year = {2011}
}

@article{Chitre2012,
  author = {Chitre, Mandar and Kuselan, Subash and Pallayil, Venugopalan},
  doi = {10.1121/1.4733553},
  issn = {1520-8524},
  journal = {The Journal of the Acoustical Society of America},
  month = {aug},
  number = {2},
  pages = {838--47},
  pmid = {22894207},
  title = {{Ambient noise imaging in warm shallow waters; robust statistical algorithms and range estimation.}},
  url = {https://doi.org/10.1121/1.4733553},
  volume = {132},
  year = {2012}
}

@article{Vishnu2024,
  author = {Vishnu, Hari and Soorya, V. R. and Chitre, Mandar and Too, Yuen Min and Koay, Teong Beng and Ho, Abel},
  doi = {10.1016/j.marenvres.2024.106571},
  isbn = {0000000324700},
  issn = {18790291},
  journal = {Marine Environmental Research},
  title = {{Machine-learning based detection of marine mammal vocalizations in snapping-shrimp dominated ambient noise}},
  url = {https://doi.org/10.1016/j.marenvres.2024.106571},
  volume = {199},
  year = {2024}
}

@article{Kennedy2010,
  author = {Kennedy, E. V. and Holderied, M. W. and Mair, J. M. and Guzman, H. M. and Simpson, S. D.},
  journal = {Journal of Experimental Marine Biology and Ecology},
  doi = {10.1016/j.jembe.2010.08.017},
  number = {1-2},
  pages = {85--92},
  title = {{Spatial patterns in reef-generated noise relate to habitats and communities: Evidence from a Panamanian case study}},
  url = {https://doi.org/10.1016/j.jembe.2010.08.017},
  volume = {395},
  year = {2010}
}

@article{Sueur2014,
  author = {Sueur, J{\'{e}}r{\^{o}}me and Farina, Almo and Gasc, Amandine and Pieretti, Nadia and Pavoine, Sandrine},
  doi = {10.3813/AAA.918757},
  journal = {Acta Acustica united with Acustica},
  number = {4},
  pages = {772--781},
  title = {{Acoustic indices for biodiversity assessment and landscape investigation}},
  url = {https://doi.org/10.3813/AAA.918757},
  volume = {100},
  year = {2014}
}

@article{JOHNSON1947,
  author = {Johnson, Martin W. and Everest, F. Alton and Young, Robert W.},
  doi = {10.2307/1538284},
  issn = {0006-3185},
  journal = {The Biological Bulletin},
  month = {oct},
  number = {2},
  pages = {122--138},
  title = {{The role of Snapping Shrimp (Crangon and Synalpheus) in the production of underwater noise in the sea}},
  url = {https://doi.org/10.2307/1538284},
  volume = {93},
  year = {1947}
}

@article{McCammon2025,
  author = {McCammon, Seth and Formel, Nathan and Jarriel, Sierra and Mooney, T. Aran},
  title = {Rapid detection of fish calls within diverse coral reef soundscapes using a convolutional neural networka)},
  journal = {The Journal of the Acoustical Society of America},
  volume = {157},
  number = {3},
  pages = {1665-1683},
  year = {2025},
  month = {03},
  issn = {0001-4966},
  doi = {10.1121/10.0035829},
  url = {https://doi.org/10.1121/10.0035829},
  eprint = {https://pubs.aip.org/asa/jasa/article-pdf/157/3/1665/20431869/1665\_1\_10.0035829.pdf},
}

@article{Erbe2019,
  author = {Erbe, Christine and Marley, Sarah A. and Schoeman, Ren{\'{e}}e P. and Smith, Joshua N. and Trigg, Leah E. and Embling, Clare Beth},
  doi = {10.3389/fmars.2019.00606},
  issn = {22967745},
  journal = {Frontiers in Marine Science},
  number = {October},
  title = {The Effects of Ship Noise on Marine Mammals—A Review},
  url = {https://doi.org/10.3389/fmars.2019.00606},
  volume = {6},
  year = {2019}
}

@article{Low2019,
  author = {Low, Jeffrey K.Y. and Fong, Jenny and Todd, Peter A. and Chou, Loke Ming and Bauman, Andrew G.},
  doi = {10.1111/jpy.12818},
  issn = {15298817},
  journal = {Journal of Phycology},
  number = {2},
  pages = {289--296},
  pmid = {30506680},
  title = {{Seasonal variation of Sargassum ilicifolium (Phaeophyceae) growth on equatorial coral reefs}},
  url = {https://doi.org/10.1111/jpy.12818},
  volume = {55},
  year = {2019}
}

@article{duarte,
author = {Carlos M. Duarte and Lucille Chapuis and Shaun P. Collin and Daniel P. Costa and Reny P. Devassy and Victor M. Eguiluz and Christine Erbe and Timothy A. C. Gordon and Benjamin S. Halpern and Harry R. Harding and Michelle N. Havlik and Mark Meekan and Nathan D. Merchant and Jennifer L. Miksis-Olds and Miles Parsons and Milica Predragovic and Andrew N. Radford and Craig A. Radford and Stephen D. Simpson and Hans Slabbekoorn and Erica Staaterman and Ilse C. Van Opzeeland and Jana Winderen and Xiangliang Zhang and Francis Juanes },
title = {The soundscape of the {Anthropocene} ocean},
journal = {Science},
volume = {371},
number = {6529},
pages = {eaba4658},
year = {2021},
doi = {10.1126/science.aba4658},
url = {https://doi.org/10.1126/science.aba4658},
eprint = {https://www.science.org/doi/pdf/10.1126/science.aba4658}}

@article{Raick2025,
author = {Raick, X and Parmentier, E and Lecchini, D and Gervaise, C and Bertucci, F and Iwankow, G and Chancerelle, Y and Siu, G and {Di Iorio}, L},
journal = {Royal Society Open Science},
publisher = {The Royal Society},
title = {{Highlighting the resilience potential of marine protected areas in the face of coral bleaching with passive acoustic monitoring}},
year = {2025},
doi = {10.1098/rsos.241938},
url = {https://doi.org/10.1098/rsos.241938}
}

@electronic{Fishsounds,
	author = {Looby, A and Vela, S and Cox, K and Riera, A and Davies, HL and Spriel, B and Murchy, K and Bravo, S and Rountree, R and Juanes, F and Reynolds, LK and Martin CW},
	title = {{FishSounds}},
	url = {{http://www.fishsounds.net}},
	year = {2026},
    note = {[Online; accessed 8-June-2026]}
}

@article{Hodson2025,
author = {Hodson, Emma Jayne and Cox, Kieran and Juanes, Francis and Looby, Audrey},
doi = {10.1111/jfb.16030},
issn = {10958649},
journal = {Journal of Fish Biology},
number = {4},
pages = {990--995},
pmid = {39681114},
title = {{Actively soniferous tropical reef fishes are diverse, vulnerable, and valuable}},
volume = {106},
year = {2025},
url = {https://doi.org/10.1111/jfb.16030}
}

@article{Chan2024,
author = {Chan, Y.K. Samuel and Ng, C.S. Lionel and Tun, Karenne P.P. and Chou, Loke Ming and Huang, Danwei},
doi = {10.1016/j.ecolind.2024.112143},
issn = {1470160X},
journal = {Ecological Indicators},
month = {jul},
pages = {112143},
title = {{Decadal decline of functional diversity despite increasing taxonomic and phylogenetic diversity of coral reefs under chronic urbanisation stress}},
url = {https://doi.org/10.1016/j.ecolind.2024.112143},
volume = {164},
year = {2024}
}

@incollection{ng2018coral,
  title={Coral reef restoration in Singapore—Past, present and future},
  author={Ng, Chin-Soon Lionel and Chou, Loke Ming},
  booktitle={Sustainability Matters: Environmental Management in the Anthropocene},
  pages={3--23},
  year={2018},
  publisher={World Scientific},
  doi = {10.1142/9789813230620_0001},
  url = {https://doi.org/10.1142/9789813230620\_0001}
}

@article{chou1996response,
  title={Response of {Singapore} reefs to land reclamation},
  author={Chou, LM},
  journal={Galaxea},
  volume={13},
  number={1},
  pages={85--92},
  year={1996},
  url = {https://coralreef.nus.edu.sg/publications/Chou1996Galaxea.pdf}
}

@article{obura2019coral,
  title={Coral reef monitoring, reef assessment technologies, and ecosystem-based management},
  author={Obura, David O and Aeby, Greta and Amornthammarong, Natchanon and Appeltans, Ward and Bax, Nicholas and Bishop, Joe and Brainard, Russell E and Chan, Samuel and Fletcher, Pamela and Gordon, Timothy AC and others},
  journal={Frontiers in Marine Science},
  volume={6},
  pages={580},
  year={2019},
  publisher={Frontiers Media SA},
  doi = {10.3389/fmars.2019.00580},
  url = {https://doi.org/10.3389/fmars.2019.00580}
}

@article{jaap2000coral,
  title={Coral reef restoration},
  author={Jaap, Walter C},
  journal={Ecological engineering},
  volume={15},
  number={3-4},
  pages={345--364},
  year={2000},
  publisher={Elsevier},
  doi = {10.1016/S0925-8574(00)00085-9},
  url = {https://doi.org/10.1016/S0925-8574(00)00085-9}
}

@incollection{chitre2012snapping,
  title={Snapping shrimp dominated natural soundscape in {Singapore} waters},
  author={Chitre, Mandar and Legg, Matthew and Koay, Teong-Beng},
  booktitle={Contrib Mar Sci},  
  address   = {Singapore},
  volume={2012},
  pages={127--134},
  year={2012}
}

@article{lobel1992sounds,
  title={Sounds produced by spawning fishes},
  author={Lobel, Phillip S},
  journal={Environmental Biology of Fishes},
  volume={33},
  number={4},
  pages={351--358},
  year={1992},
  publisher={Springer},
  doi = {10.1007/BF00010947},
  url = {https://doi.org/10.1007/BF00010947}
}

@article{lammers2008ecological,
  title={An ecological acoustic recorder (EAR) for long-term monitoring of biological and anthropogenic sounds on coral reefs and other marine habitats},
  author={Lammers, Marc O and Brainard, Russell E and Au, Whitlow WL and Mooney, T Aran and Wong, Kevin B},
  journal={The Journal of the Acoustical Society of America},
  volume={123},
  number={3},
  pages={1720--1728},
  year={2008},
  publisher={Acoustical Society of America},
  doi = {10.1121/1.2836780},
  url = {https://doi.org/10.1121/1.2836780}
}

@phdthesis{mcwilliam2018coral,
  title={Coral reef soundscapes: The use of passive acoustic monitoring for long-term ecological survey},
  author={McWilliam, Jamie Neish},
  year={2018},
  school={Curtin University}
}

@article{nedelec2015soundscapes,
  title={Soundscapes and living communities in coral reefs: temporal and spatial variation},
  author={Nedelec, Sophie L and Simpson, Stephen D and Holderied, Marc and Radford, Andrew N and Lecellier, Gael and Radford, Craig and Lecchini, David},
  journal={Marine Ecology Progress Series},
  volume={524},
  pages={125--135},
  year={2015},
  doi = {10.3354/meps11175},
  url = {https://doi.org/10.3354/meps11175}
}

@article{Lin2023,
author = {Lin, Tzu Hao and Sinniger, Frederic and Harii, Saki and Akamatsu, Tomonari},
doi = {10.5670/oceanog.2023.s1.7},
issn = {10428275},
journal = {Oceanography},
number = {1},
pages = {20--27},
title = {{Using Soundscapes to Assess Changes in Coral Reef Social-Ecological Systems}},
volume = {36},
year = {2023},
url = {https://doi.org/10.5670/oceanog.2023.s1.7}
}

@article{Siddagangaiah2022,
author = {Siddagangaiah, Shashidhar and Chen, Chi Fang and Hu, Wei Chun and Farina, Almo},
doi = {10.1038/s43247-022-00442-5},
isbn = {4324702200442},
issn = {26624435},
journal = {Communications Earth and Environment},
number = {1},
publisher = {Springer US},
title = {{The dynamical complexity of seasonal soundscapes is governed by fish chorusing}},
volume = {3},
year = {2022},
url = {https://doi.org/10.1038/s43247-022-00442-5}
}

@article{Lin2021,
author = {Lin, Tzu Hao and Akamatsu, Tomonari and Sinniger, Frederic and Harii, Saki},
doi = {10.1016/j.biocon.2020.108901},
issn = {00063207},
journal = {Biological Conservation},
number = {December 2020},
pages = {108901},
publisher = {Elsevier Ltd},
title = {{Exploring coral reef biodiversity via underwater soundscapes}},
url = {https://doi.org/10.1016/j.biocon.2020.108901},
volume = {253},
year = {2021}
}

@inproceedings{Stoller2018,
archivePrefix = {arXiv},
arxivId = {1806.03185},
author = {Stoller, Daniel and Ewert, Sebastian and Dixon, Simon},
booktitle = {Proceedings of the 19th International Society for Music Information Retrieval Conference, ISMIR 2018},
doi = {10.48550/arXiv.1806.03185},
eprint = {1806.03185},
isbn = {9782954035123},
pages = {334--340},
title = {{Wave-U-Net: A multi-scale neural network for end-to-end audio source separation}},
year = {2018},
url = {https://doi.org/10.48550/arXiv.1806.03185}
}

@article{Defossez2021,
archivePrefix = {arXiv},
arxivId = {2111.03600},
author = {D{\'{e}}fossez, Alexandre},
doi = {10.48550/arXiv.2111.03600},
eprint = {2111.03600},
journal = {arXiv preprint},
pages = {1--13},
title = {{Hybrid Spectrogram and Waveform Source Separation}},
url = {https://doi.org/10.48550/arXiv.2111.03600},
year = {2021}
}

@article{Fleishman2023,
author = {Fleishman, Erica and Cholewiak, Danielle and Gillespie, Douglas and Helble, Tyler and Klinck, Holger and Nosal, Eva‐Marie and Roch, Marie A.},
doi = {10.1111/brv.12969},
issn = {1464-7931},
journal = {Biological Reviews},
month = {oct},
number = {5},
pages = {1633--1647},
title = {{Ecological inferences about marine mammals from passive acoustic data}},
url = {https://doi.org/10.1111/brv.12969},
volume = {98},
year = {2023}
}

@techreport{Tyack2023,
author = {Tyack, P L and Akamatsu, T and Boebel, O and Chapuis, L and Debusschere, E and {De Jong}, C and Erbe, C and Evans, K and Gedamke, J and Gridley, T and Haralabus, G. and Jenkins, R. and Miksis-Olds, J. and Sagen, H. and Thomsen, F. and Thomisch, K. and Urban, E.},
booktitle = {Ocean Sound Essential Ocean Variable Implementation Plan. International Quiet Ocean Experiment},
doi = {10.5281/zenodo.10067187},
institution = {International Quiet Ocean Experiment, Scientific Committee on Oceanic Research and Partnership for Observation of the Global Ocean},
pages = {87},
title = {{Ocean Sound Essential Ocean Variable Implementation Plan.}},
url = {https://doi.org/10.5281/zenodo.10067187},
year = {2023}
}

@article{Kinda2013,
author = {Kinda, G. Bazile and Simard, Yvan and Gervaise, C{\'{e}}dric and Mars, J{\'{e}}rome I. and Fortier, Louis},
doi = {10.1121/1.4808330},
issn = {0001-4966},
journal = {The Journal of the Acoustical Society of America},
month = {jul},
number = {1},
pages = {77--87},
title = {{Under-ice ambient noise in Eastern Beaufort Sea, Canadian Arctic, and its relation to environmental forcing}},
url = {https://doi.org/10.1121/1.4808330},
volume = {134},
year = {2013}
}

@article{Mahmood2017,
	author = {Mahmood, A. and Chitre, M. and Vishnu, H.},
	doi = {10.1109/JOE.2017.2731058},
	issn = {03649059},
	journal = {IEEE Journal of Oceanic Engineering},
	number = {4},
	title = {{Locally optimal inspired detection in snapping shrimp noise}},
	volume = {42},
	year = {2017},
	url = {https://doi.org/10.1109/JOE.2017.2731058}
}

@article{Raick2025a,
author = {Raick, Xavier and Gervaise, C{\'{e}}dric and Lecchini, David and Parmentier, {\'{E}}ric},
doi = {10.1080/09524622.2025.2474482},
issn = {21650586},
journal = {Bioacoustics},
number = {3},
pages = {207--220},
publisher = {Taylor & Francis},
title = {{Limitations of $\alpha$-acoustic diversity indices in assessing invertebrate sounds in coral reefs}},
url = {https://doi.org/10.1080/09524622.2025.2474482},
volume = {34},
year = {2025}
}

@article{freeman2016rapidly,
  title={Rapidly obtained ecosystem indicators from coral reef soundscapes},
  author={Freeman, Lauren A and Freeman, Simon E},
  journal={Marine Ecology Progress Series},
  volume={561},
  pages={69--82},
  year={2016},
  doi={10.3354/meps11938},
  url={https://doi.org/10.3354/meps11938}
}

@article{kaplan2015coral,
  title={Coral reef species assemblages are associated with ambient soundscapes},
  author={Kaplan, Maxwell B and Mooney, T Aran and Partan, Jim and Solow, Andrew R},
  journal={Marine Ecology Progress Series},
  volume={533},
  pages={93--107},
  year={2015},
  doi={10.3354/meps11382},
  url={https://doi.org/10.3354/meps11382}
}

@article{harris2016ecoacoustic,
  title={Ecoacoustic indices as proxies for biodiversity on temperate reefs},
  author={Harris, Sydney A and Shears, Nick T and Radford, Craig A},
  journal={Methods in Ecology and Evolution},
  volume={7},
  number={6},
  pages={713--724},
  year={2016},
  publisher={Wiley Online Library},
  doi={10.1111/2041-210x.12527},
  url={https://doi.org/10.1111/2041-210x.12527}
}

@article{bertucci2016acoustic,
  title={Acoustic indices provide information on the status of coral reefs: an example from {Moorea} Island in the {South Pacific}},
  author={Bertucci, Fr{\'e}d{\'e}ric and Parmentier, Eric and Lecellier, Ga{\"e}l and Hawkins, Anthony D and Lecchini, David},
  journal={Scientific Reports},
  volume={6},
  number={1},
  pages={1--9},
  year={2016},
  publisher={Nature Publishing Group},
  doi={10.1038/srep33326},
  url={https://doi.org/10.1038/srep33326}
}

@article{kaplan2018acoustic,
  title={Acoustic and biological trends on coral reefs off {Maui, Hawaii}},
  author={Kaplan, Maxwell B and Lammers, Marc O and Zang, Eden and Mooney, T Aran},
  journal={Coral Reefs},
  volume={37},
  number={1},
  pages={121--133},
  year={2018},
  publisher={Springer},
  doi={10.1007/s00338-017-1638-x},
  url={https://doi.org/10.1007/s00338-017-1638-x}
}

@article{versluis2000snapping,
  title={How snapping shrimp snap: through cavitating bubbles},
  author={Versluis, Michel and Schmitz, Barbara and Von der Heydt, Anna and Lohse, Detlef},
  journal={Science},
  volume={289},
  number={5487},
  pages={2114--2117},
  year={2000},
  publisher={American Association for the Advancement of Science},
  doi={10.1126/science.289.5487.2114},
  url={https://doi.org/10.1126/science.289.5487.2114}
}

@article{lin2017improving,
  title={Improving biodiversity assessment via unsupervised separation of biological sounds from long-duration recordings},
  author={Lin, Tzu-Hao and Fang, Shih-Hua and Tsao, Yu},
  journal={Scientific reports},
  volume={7},
  number={1},
  pages={4547},
  year={2017},
  publisher={Nature Publishing Group UK London},
  doi={10.1038/s41598-017-04790-7},
  url={https://doi.org/10.1038/s41598-017-04790-7}
}

@article{elise2019optimised,
  title={An optimised passive acoustic sampling scheme to discriminate among coral reefs’ ecological states},
  author={Elise, Simon and Bailly, Arthur and Urbina-Barreto, Isabel and Mou-Tham, G{\'e}rard and Chiroleu, Fr{\'e}d{\'e}ric and Vigliola, Laurent and Robbins, William D and Bruggemann, J Henrich},
  journal={Ecological Indicators},
  volume={107},
  pages={105627},
  year={2019},
  publisher={Elsevier},
  doi={10.1016/j.ecolind.2019.105627},
  url={https://doi.org/10.1016/j.ecolind.2019.105627}
}

@article{williams2022enhancing,
  title={Enhancing automated analysis of marine soundscapes using ecoacoustic indices and machine learning},
  author={Williams, Ben and Lamont, Timothy AC and Chapuis, Lucille and Harding, Harry R and May, Eleanor B and Prasetya, Mochyudho E and Seraphim, Marie J and Jompa, Jamaluddin and Smith, David J and Janetski, Noel and others},
  journal={Ecological Indicators},
  volume={140},
  pages={108986},
  year={2022},
  publisher={Elsevier},
  doi={10.1016/j.ecolind.2022.108986},
  url={https://doi.org/10.1016/j.ecolind.2022.108986}
}

@inproceedings{potter1997acoustic,
  title={Acoustic imaging \& the natural soundscape in {Singapore} waters},
  author={Potter, John R and Wei, Lim Tze and Chitre, Mandar},
  booktitle={Proceedings of Mindef-NUS joint seminar},
  pages={141--147},
  year={1997},
  organization={Citeseer}
}

@inproceedings{Potter1997,
  author    = {J. R. Potter and T. W. Lim and M. Chitre},
  title     = {Ambient Noise Environments in Shallow Tropical Seas 
               and the Implications for Acoustic Sensing},
  booktitle = {Proc. Oceanology International '97 Pacific Rim},
  address   = {Singapore},
  year      = {1997},
  pages     = {191--199},
  note      = {[Online]. Available: 
               \url{https://arl.nus.edu.sg/wp-content/publications/Potter1997OceanologyIntl.pdf}}
}

@manual{tidetables2021,
  title        = {Singapore Tide Tables Year 2021},
  organization = {Hydrographic Division, Maritime and Port Authority 
                  of Singapore},
  address      = {Singapore},
  year         = {2020},
  note         = {ISSN: 0129-0282},
  url = {https://www.mpa.gov.sg/who-we-are/newsroom-resources/publications/singapore-tide-tables}
}

@article{yan2006background,
  title={Background noise cancellation of manatee vocalizations using an adaptive line enhancer},
  author={Yan, Zheng and Niezrecki, Christopher and Cattafesta, Louis N and Beusse, Diedrich O},
  journal={The Journal of the Acoustical Society of America},
  volume={120},
  number={1},
  pages={145--152},
  year={2006},
  publisher={AIP Publishing},
  doi={10.1121/1.2202885},
  url={https://doi.org/10.1121/1.2202885}
}

@article{luo2019conv,
  title={Conv-tasnet: Surpassing ideal time--frequency magnitude masking for speech separation},
  author={Luo, Yi and Mesgarani, Nima},
  journal={IEEE/ACM transactions on audio, speech, and language processing},
  volume={27},
  number={8},
  pages={1256--1266},
  year={2019},
  url = {https://doi.org/10.1109/TASLP.2019.2915167},
  publisher={IEEE},
  doi={10.1109/TASLP.2019.2915167}
}

@article{looby2023fishsounds,
  title={FishSounds Version 1.0: A website for the compilation of fish sound production information and recordings},
  author={Looby, Audrey and Vela, Sarah and Cox, Kieran and Riera, Amalis and Bravo, Santiago and Davies, Hailey L and Rountree, Rodney and Reynolds, Laura K and Martin, Charles W and Matwin, Stan and others},
  journal={Ecological Informatics},
  volume={74},
  pages={101953},
  year={2023},
  publisher={Elsevier},
  doi={10.1016/j.ecoinf.2023.101953},
  url={https://doi.org/10.1016/j.ecoinf.2023.101953}
}

@article{irfan2021deepship,
  title={DeepShip: An underwater acoustic benchmark dataset and a separable convolution based autoencoder for classification},
  author={Irfan, Muhammad and Jiangbin, ZHENG and Ali, Shahid and Iqbal, Muhammad and Masood, Zafar and Hamid, Umar},
  journal={Expert Systems with Applications},
  volume={183},
  pages={115270},
  year={2021},
  publisher={Elsevier},
  doi={10.1016/j.eswa.2021.115270},
  url={https://doi.org/10.1016/j.eswa.2021.115270}
}

@article{pieretti2011new,
  title={A new methodology to infer the singing activity of an avian community: The Acoustic Complexity Index (ACI)},
  author={Pieretti, Nadia and Farina, Almo and Morri, Davide},
  journal={Ecological indicators},
  volume={11},
  number={3},
  pages={868--873},
  year={2011},
  publisher={Elsevier},
  doi={10.1016/j.ecolind.2010.11.005},
  url={https://doi.org/10.1016/j.ecolind.2010.11.005}
}

@article{coker2012interactive,
  title={Interactive effects of live coral and structural complexity on the recruitment of reef fishes},
  author={Coker, DJ and Graham, NAJ and Pratchett, MS},
  journal={Coral reefs},
  volume={31},
  pages={919--927},
  year={2012},
  publisher={Springer},
  doi={10.1007/s00338-012-0920-1},
  url={https://doi.org/10.1007/s00338-012-0920-1}
}

@article{lamont2022sound,
  title={The sound of recovery: Coral reef restoration success is detectable in the soundscape},
  author={Lamont, Timothy AC and Williams, Ben and Chapuis, Lucille and Prasetya, Mochyudho E and Seraphim, Marie J and Harding, Harry R and May, Eleanor B and Janetski, Noel and Jompa, Jamaluddin and Smith, David J and others},
  journal={Journal of Applied Ecology},
  volume={59},
  number={3},
  pages={742--756},
  year={2022},
  publisher={Wiley Online Library},
  doi={10.1111/1365-2664.14089},
  url={https://doi.org/10.1111/1365-2664.14089}
}

\begin{IEEEbiography}[{\includegraphics[width=1in,height=1.25in,clip,keepaspectratio]{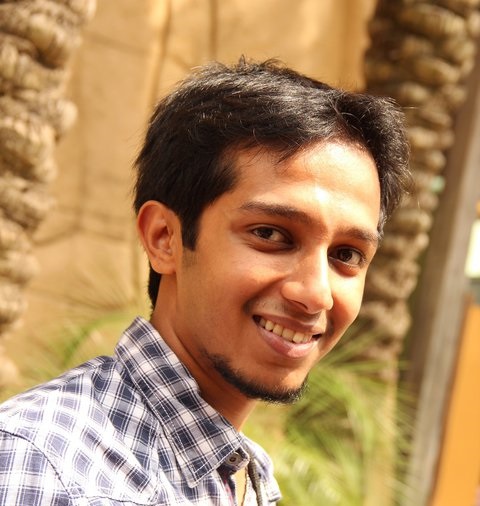}}]{Hari Vishnu}(S’05–M’13–SM’20) is a Senior Research Fellow at the Acoustic Research Laboratory, National University of Singapore. His interests include machine learning for underwater applications, bio-acoustics and processing in impulsive noise. These are used in a wide range of underwater applications ranging from biodiversity or defense-related scenarios in shallow tropical waters infested with snapping shrimp noise, to polar glaciers where melt noise dominates the soundscape. From 2019, Hari is focusing on glacier cryo-acoustics, bio-acoustics based biodiversity assessment, and distributed acoustic sensing. He obtained his Ph.D from Nanyang Technological University, Singapore, in Computer Engineering on underwater signal processing including robust detection and localization. 

He is the Chief Editor on the IEEE OES Science outreach magazine Earthzine and serves on the IEEE OES Executive committee as Deputy Secretary. He was publicity chair for AUV symposium 2022 and tutorial chair for OCEANS 2024 Singapore. In 2019, he was awarded the IEEE OES YP-BOOST award which aims to encourage young professionals to participate in the society leadership. 
\end{IEEEbiography}

\begin{IEEEbiography}[{\includegraphics[width=1in,height=1.25in,clip,keepaspectratio]{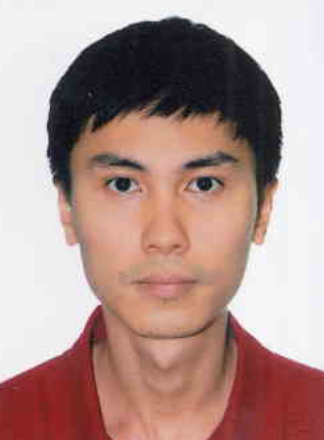}}]{Yuen Min Too}(M’15) received the Bachelor of Engineering degree in Biomedical from the Universiti Teknologi Malaysia (UTM), Malaysia in 2010, and the Ph.D. degree in underwater acoustics and signal processing from the Department of Electrical and Computer Engineering, National University of Singapore (NUS), Singapore in 2017.

From 2015 to 2025, he was with the Acoustic Research Laboratory at NUS, where he served as a Research Engineer and later as a Research Fellow. His research focused on underwater acoustics and signal processing, with particular interest in compressed sensing and array signal processing in complex underwater noise environments. He also contributed to the development of machine learning techniques for underwater sound event detection and acoustic signal denoising.

After a decade in academia, he transitioned to industry to pursue a more practical role, motivated by the desire to see his research applied in real-world settings. In 2025, he joined Subnero Pte. Ltd. as a Software Engineer, developing advanced solutions for underwater wireless communication networks.
\end{IEEEbiography}

\begin{IEEEbiography}[{\includegraphics[width=1in,height=1.25in,clip,keepaspectratio]{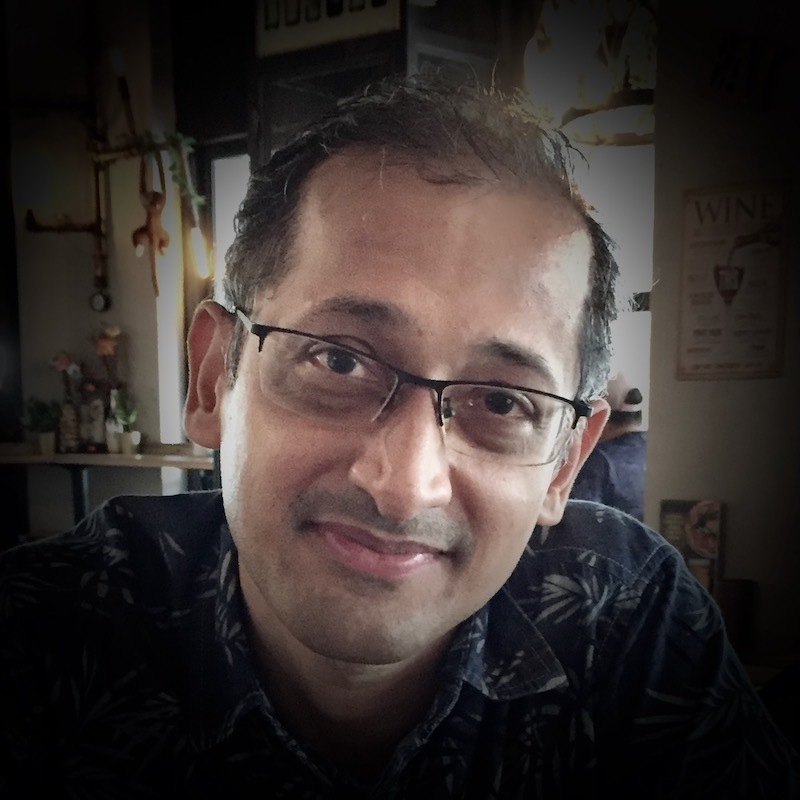}}]{Mandar Chitre}(S’04–M’05–SM’11–F'25) received the B.Eng. and M.Eng. degrees in electrical engineering from the National University of Singapore (NUS), Singapore, in 1997 and 2000, respectively, the M.Sc.
degree in bioinformatics from the Nanyang Technological University (NTU), Singapore, in 2004, and the Ph.D. degree in underwater communications from NUS, in 2006.

From 1997 to 1998, he was with the Acoustic Research Laboratory (ARL), NUS. From 1998 to
2002, he headed the technology division of a regional telecommunications solutions company.
In 2003, he rejoined ARL, initially as the Deputy Head (Research) and is currently the Head of the laboratory. He also holds a joint appointment with the Department of Electrical and Computer
Engineering, NUS as an Associate Professor. His current research interests include underwater communications, acoustic propagation modeling, marine robotics, model-based signal processing, and machine learning.

Dr. Chitre served as the Editor-in-Chief for the IEEE JOURNAL OF OCEANIC ENGINEERING from 2018 to 2023. He was the Technical Co-Chair for IEEE OCEANS 2020 Singapore – U.S. Gulf Coast and the Technical Chair for IEEE OCEANS 2024 Singapore. He was the Chairman of the student poster committee for the IEEE OCEANS 2006 Singapore and the founding Chairman for the IEEE Singapore AUV Challenge in 2013.
He was awarded the Distinguished Technical Achievement Award by the IEEE Oceanic Engineering Society in 2020 for his work on underwater communications \& networking.
\end{IEEEbiography}

\begin{IEEEbiography}[{\includegraphics[width=1in,height=1.25in,clip,keepaspectratio]{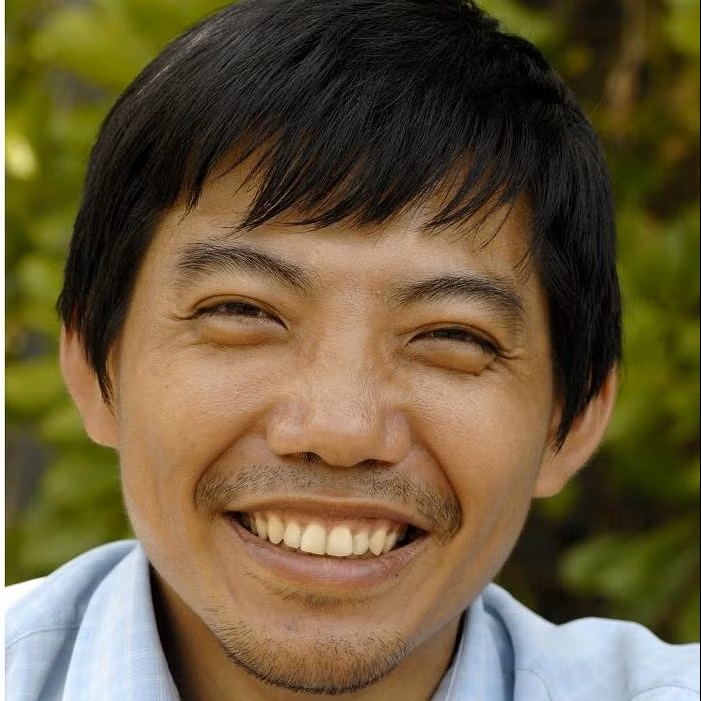}}]{Koay Teong Beng} (M’06) received the B.Eng. degree from the Universiti Teknologi Malaysia, Johor, Malaysia, in 1997, and the M.Eng. degree from the National University of Singapore, Singapore, in 2004. 
He joined the Acoustic Research Laboratory (ARL), National University of Singapore as a Research Engineer and led the development of marine autonomous platforms supporting field research in underwater acoustics, bioacoustics, underwater communications, and others.Hewas a Research Associate and a Research Fellow from 2005 to 2016 and focused on marine robotics in the later years. He is currently a Senior Engineer and Principal investigator in ARL, leading few projects in marine mammal monitoring, field robotics, and environmental sensing. His current research interests include bringing heterogeneous robots, sensors, and associated intelligence into daily environment sensing.
Mr. Beng has served in various committee positions in IEEE OES Chapter in Singapore, and as the Finance Chair for OCEANS 2020 Singapore and OCEANS 2024 Singapore.
\end{IEEEbiography}

\begin{IEEEbiography}[{\includegraphics[width=1in,height=1.25in,clip,keepaspectratio]{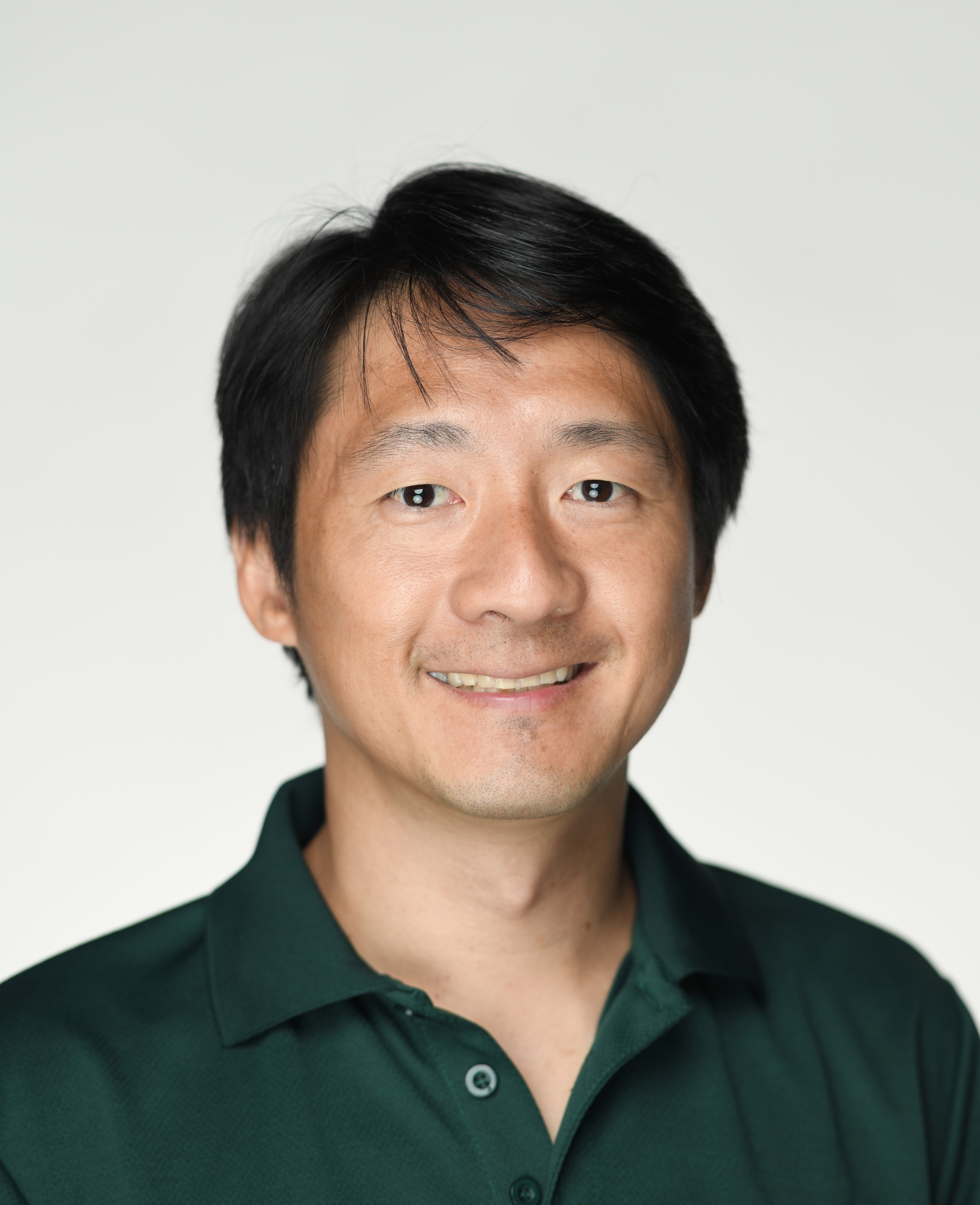}}]{Danwei Huang} is an Associate Professor at the Lee Kong Chian Natural History Museum, National University of Singapore. His research integrates the fields of ecology, evolutionary biology and marine conservation, using the coral reef as a model system to address questions on biodiversity. The questions his team is addressing are motivated by the fact that coral reefs are one of the richest and yet most threatened ecosystems on Earth, so he seeks to gain understanding of the origins, maintenance, and loss of marine biodiversity at the local, regional and global scales.
\end{IEEEbiography}

\begin{IEEEbiography}[{\includegraphics[width=1in,height=1.25in,clip,keepaspectratio]{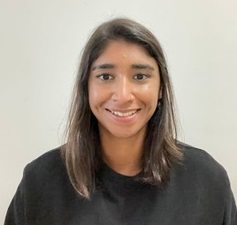}}]{Sudhanshi Jain} is a molecular and marine biologist whose earlier research focused on coral reef health, symbiont dynamics, and biodiversity across tropical marine ecosystems. During her time at the National University of Singapore’s Reef Ecology Laboratory, she led projects integrating molecular techniques—including next‑generation sequencing, qPCR, and molecular barcoding—with field‑based ecological work to investigate coral stress responses and Symbiodiniaceae community structure. Her field experience spans Singapore, Indonesia, the Red Sea, and the broader Indo‑Pacific, where she conducted coral surveys, specimen collections, and ecological assessments. Sudhanshi’s research bridged molecular biology and marine ecology to advance understanding of reef resilience in rapidly changing coastal environments.
\end{IEEEbiography}
\end{document}